# On resolving meso-scale calculations of pore-collapse-generated hotspots in energetic crystals for consistency with atomistic models

[1]Chukwudubem Okafor, [1]Jacob Herrin, [3]Catalin R. Picu, [2]Tommy Sewell, [4]John Brennan, [4]James Larentzos, and [1*]H. S. Udaykumar

[1]Department of Mechanical and Industrial Engineering, The University of Iowa, Iowa City, IA 52242

[2]Department of Chemistry and Materials Science & Engineering Institute, University of Missouri, Columbia, MO 65211

[3]Department of Mechanical, Aerospace and Nuclear Engineering, Rensselaer Polytechnic Institute, Troy, NY 12180

[4]U.S. Army Combat Capabilities Development Command (DEVCOM) Army Research Laboratory, Aberdeen Proving Ground, Maryland 21005, USA

*Corresponding author email: hs-kumar@uiowa.edu

**Abstract**

Meso-scale calculations of pore collapse and hotspot formation in energetic crystals provide closure models to macro-scale hydrocodes for predicting the shock sensitivity of energetic materials. To this end, previous works [1, 2] obtained atomistics-consistent material models for two common energetic crystals, HMX and RDX, such that pore collapse calculations adhered closely to molecular dynamics (MD) results on key features of energy localization, particularly the profiles of the collapsing pores, appearance of shear bands, and the transition from viscoplastic to hydrodynamic collapse. However, some important aspects such as the temperature distributions in the hotspot were not as well captured. One potential issue was noted but not resolved adequately in those works, namely the grid resolution that should be employed in the meso-scale calculations for various pore sizes and shock strengths. Conventional computational mechanics guidelines for selecting meshes as fine as possible, balancing computational effort, accuracy and grid independence, were shown not to produce physically consistent features associated with shear localization. Here, we examine the physics of pore collapse, shear band evolution and structure, and hotspot formation, for both HMX and RDX; we then evaluate under what conditions atomistics-consistent models yield "physically correct" (considering MD as "ground truth") hotspots for a range of pore diameters, from *nm* to microns, and for a wide range of shock strengths. The study provides insights into the effects of pore size and shock strength on pore collapse and hotspots, identifying aspects such as size-independent behaviors, and proportion of energy contained in shear as opposed to jet impact-heated regions of the hotspot. Areas for further improvement of atomistics-consistent material models are also indicated.

## 1. INTRODUCTION

Key physical phenomena that contribute to the sensitivity of EM occur at length scales ranging from molecular to macro-scale. Therefore, multi-scale modeling frameworks are necessary for understanding the shock-to-detonation transition (SDT) of energetic materials (EM) [3, 4]. In hierarchal multi-scale modeling [5], information can be extracted from smaller scales and transferred to larger scales through closure models. This strategy places the onus on obtaining physically accurate closure models at each scale in the hierarchal multi-scale model, from atomistic, through meso- and on to the macro-scale (where predictions and validations against experiments [6-8] are made). In previous works for two energetic species, HMX and RDX [1, 2], we showed that material models parameterized from atomistic simulation data can be provided as input to meso-scale continuum models. Pore collapse simulations using these MD-informed models showed general agreement with MD, for various aspects of the pore collapse and hotspot generation. However, despite the promise of the atomistics-consistent models presented in those works, we observed deviations of the continuum solutions from the corresponding MD calculations, which were attributed to "super-resolution" in the continuum simulations. This problem arises because the continuum governing equations do not possess an intrinsic physics-based cutoff length scale. Therefore, guided by conventional practice for determining grid resolution in continuum (finite difference/volume/element) approaches, each case was resolved to provide near-grid-independent solutions [9, 10]. Essentially, this amounted to very fine resolutions, for example grid sizes of 0.1 *nm* size for nanometer-sized voids, i.e., resolution finer than the molecular sizes of typical EM, such as HMX and RDX. The resolution also varied across pore sizes, guided by the heuristic criterion of about ~500 mesh points across a pore diameter for grid-independence [9, 10]. The overall effect of these choices on the physical correctness of the shock-pore-hotspot dynamics was unclear. In this work, we address these issues to determine the most effective, computationally economical, and physically accurate way to set up meso-scale calculations of energy localization in EM.

Determining whether continuum calculations of hotspots provide "accurate" temperature distributions is challenging. Experimental measurements of temperatures are few [11-14] and are hampered by issues including spatiotemporal resolution [15, 16], calibration [17], and even fundamental questions regarding whether the measured value is the "temperature" given the highly non-equilibrium state of the material in a shock-induced hotspot. On the other hand, MD calculations can provide "ground truth" but are also subject to limitations, including neglect of quantum effects [18], and uncertainties associated with potentials that underpin models such as ReaxFF variants [19, 20]. Inert MD calculations, utilizing the well-tested and reliable Smith-Bharadwaj potentials [21] offer a starting point for establishing temperatures and post-collapse distributions in hotspots. Even so, such calculations are classical in construction and therefore the specific heat of the EM is inherently constant and the Gruneisen coefficient is likewise constrained. Nevertheless, since the solid ground in the MD context is in classical inert models, in [1, 2] the pore collapse behavior and hotspots from atomistics-consistent models were evaluated against classical all-atom MD. In this work, we continue this vein of work and evaluate how to arrange continuum simulations to obtain better agreement with MD for the temperature distributions, particularly in the tail of the distribution. Once the physical fidelity of meso-scale calculations is ensured, one can exploit several routes to connect meso-scale energy localization data to make macro-scale predictions of EM sensitivity [6, 7, 22, 23].

Both accurate modeling *and* computation of shock-pore dynamics are crucial for quantitative, multi-scale, meso-informed burn model development that can lead to more precise control and design of energy release from EM. In recent work, atomistics-consistent models for RDX [1] and HMX [2] were developed to calculate pore collapse for a range of shock strengths, spanning viscoplasticity-dominated to hydrodynamics-dominated collapse. In [1] and [2] we showed that key quantitative metrics for assessing the fidelity of the continuum calculations were well captured, assessed against MD simulations. However, there were discrepancies between the MD and continuum calculations, particularly with regard to the temperature distributions in the hotspots. This is a matter of concern for quantitative predictive modeling of reactive heat release in a shocked EM that may transition to detonations. The magnitude and distribution of temperatures in hotspots are of crucial importance [24, 25] since meso-scale calculations of the reaction rate progress and advance of the burn front initiated by the hotspots rely on Arrhenius chemical kinetics models [26, 27]; due to the exponential dependence on temperatures, the tails of the temperature distributions are crucial for obtaining physically accurate burn rates. To be sure, there remain uncertainties in the reaction kinetic models, including for common EM such as HMX [28] and RDX [29]. Nevertheless, for any selected reaction kinetics scheme, the pre-requisite for calculating reaction progress is the determination of physically correct temperature distributions in hotspots. The current work aims to reach closer agreement with the tandem MD data, with particular regard to hotspot temperatures, at least in the inert case where the MD calculations are more well settled.

Extending the work presented in [1] and [2] this paper addresses the calculation of key pore collapse metrics and hotspot quantities of interest for the widely used energetic crystals β-HMX and α-RDX. The extensions consist of examining a range of pore sizes as well as impact velocities (shock strengths) and comparing with MD data where available. Thus, for both materials we examine the issue of grid resolution over a wide range of pore sizes and shock strengths and quantitatively assess the ability of the models and numerical setups in the continuum calculations to yield physically correct pore collapse and hotspot behaviors. In Sec. 2, the continuum and MD models and formulations are briefly outlined; details of the governing equations, numerical approach and computations involved are available in previous works [1, 2]. Section 3 presents the results of the study, examining first the effect of grid resolution on the agreement of the continuum results with MD, followed by analysis of the characteristics of shear bands and hotspot temperatures as the pore radius is increased from nanometer to micron size, the latter end of the range pertaining to "real" hotspot sizes relevant to sensitivity of EM. Finally, in Sec. 4, we present conclusions from the study and point to directions for future investigations.

## 2. METHODS

### 2.2 Continuum Model

The conservation equations of mass, momentum and energy along with MD-derived material models for RDX [1] and HMX [2] were implemented in SCIMITAR3D [30-32], a multi-material hydrocode which employs a fixed Cartesian grid and compressible, Eulerian framework. The governing equations and numerical techniques for their solution are presented in several previous works [33-35]. The governing equations are solved to obtain density ($\rho$), velocity fields ($u_i$), internal energy ($e$) and deviatoric stress components ($S_{ij}$). The system of conservation laws is closed by providing constitutive models for the pressure, temperature, and flow rule for plastic deformation. These constitutive models applicable to both HMX and RDX materials are briefly outlined in the following.

Temperature is computed using a caloric equation of state in the following form:

$$T(\rho, e) = T_0 + \frac{e - e_c}{c_v}, \qquad (1)$$

Where $T_0$ is the reference temperature, $c_v$ is a constant, isochoric specific heat, and $e_c$ corresponds to the athermal or "cold" internal energy due to compression. In this work, to maintain strict consistency with MD, a constant $c_v$ is used for both RDX (1980 J kg$^{-1}$ K$^{-1}$) and HMX (2359 J kg$^{-1}$ K$^{-1}$).

Shock heating and plastic dissipation can lead to the melting of HMX and RDX; in the present model, once the temperature exceeds the melting point of the energetic, the deviatoric stress terms are set to and remain at zero. The HMX model for the melting temperature is taken from Kroonblawd and Austin (K-A) [36], while the RDX model is taken from Kroonblawd and Springer (K-S) [37] who used MD simulations of crystal-liquid coexistence to determine the melt curve $T_m = T_m(P)$ for $0 \leq P \leq 5$ GPa (HMX) and 30 GPa (RDX). Both melt temperature models are of Simon-Glatzel form [38],

$$T_m(P) = T_{m,ref} \left[ 1 + \frac{P - P_{ref}}{a} \right]^{\frac{1}{c}} \qquad (2)$$

where $T_{m,ref}$ and $P_{ref}$ are the reference values for melt temperature and pressure, respectively. The quantities $a$ and $c$ are fit model parameters and are given in Table I.

To model pressure in HMX and RDX, an MD-derived Mie-Gruneisen equation of state (EOS) is used in the following form:

$$p = p_c(\rho) + \frac{\Gamma}{V}[e - e_c(\rho)], \qquad (3)$$

where $V$ is specific volume, $\Gamma$ is the Grüneisen coefficient. The cold pressure $p_c$ is obtained from the Birch-Murnaghan EOS:

$$p_c = \frac{3}{2} K_0 \left[ \left( \frac{\rho_0}{\rho} \right)^{-\frac{7}{3}} - \left( \frac{\rho_0}{\rho} \right)^{-\frac{5}{3}} \right] \left[ 1 + \frac{3}{4} (K_0^{'} - 4) \left\{ \left( \frac{\rho_0}{\rho} \right)^{-\frac{2}{3}} - 1 \right\} \right], \quad (4)$$

where the coefficients $K_0$ and $K_0^{'}$ are the isothermal bulk modulus and its initial pressure derivative, respectively, and $V_0$ is the reference volume at (300 K, 1 atm). Values for $K_0$, $K_0^{'}$, and $V_0$ were determined by fitting Eq. (4) to MD predictions for the 300 K isotherm and are reported in Table I.

The Grüneisen coefficient is represented as:

$$\Gamma = \Gamma_0 + \gamma_1 \left( \frac{\rho_0}{\rho} \right)^{\beta_1} + \gamma_2 \left( \frac{\rho_0}{\rho} \right)^{\beta_2} \quad (5)$$

where $\Gamma_0$ corresponds to the zero-pressure Gruneisen parameter and model coefficients $\gamma_1, \gamma_2, \beta_1$, and $\beta_2$ are fit to MD simulation data based on parameterized Helmholtz free energies. The values of the coefficients for HMX and RDX are reported in Table I.

The shear modulus is described using a MD-derived model. For HMX, shear modulus is modeled to be dependent on pressure [39] whereas for RDX, the shear modulus is both temperature and pressure dependent [40] as follows:

$$G_{HMX}(P) = G_0 + a_1 P + a_2 P^2 \quad (6)$$

$$G_{RDX}(P,T) = G_0 + a_1 P + a_2 (T - T_0) \quad (7)$$

For both descriptions, $G_0, a_1$ and $a_2$ are obtained via bivariate first-degree polynomial fits of the Voigt-Reuss-Hill averaged elastic constants[39, 40], reported in Table I.

In a prior work [2], we advanced a MD-guided, rate-dependent, isotropic, Johnson-Cook J2 (JC-J2) plasticity model for HMX. In addition, recent work used crystal plasticity simulations to develop a similar model for RDX [40]. The model includes strain hardening, strain-rate hardening, and thermal softening:

$$\sigma_Y = [A + B \varepsilon_p^n][1 + C \, ln(1 + \dot{\varepsilon}^*)] \left[ 1 - \left( \frac{T - T_{ref}}{T_m - T_{ref}} \right)^M \right]. \quad (8)$$

Here, $\varepsilon_p$ is the equivalent plastic strain, $\dot{\varepsilon}$ is the plastic strain rate, $\dot{\varepsilon}_0$ is a reference plastic strain rate, and $\dot{\varepsilon}^* = \dot{\varepsilon}/\dot{\varepsilon}_0$; $T$ is the temperature and $T_{ref}$ is the reference temperature of the material (300 K); and $A, B, C, n,$ and $M$ are Johnson-Cook model coefficients. Values of these parameters for HMX and RDX are provided in Table I, along with relevant references.

In recent work [41], we showed that the Johnson-Cook model in Eq. (8) does not yield pore collapse behavior that reproduces MD results; this is due to the lack of strain localization in the model, which is a strong feature of pore collapse seen in the MD calculations [42, 43]. To address this shortcoming a modified Johnson-Cook (M-JC) model was proposed, guided by MD calculations of Khan and Picu [44]. The model initiates strain-softening at the surface of the pore leading to localization and the formation of shear bands that propagate away from the pore surface, yielding results for pore collapse and hotspot temperatures in general agreement with MD [1, 2]. The localization is triggered at a threshold value of the deviatoric (shear) stress at the surface of the pore. The threshold value ($\sigma_{CSSL}$) is related to the critical shear stress for localization (CSSL) and has been determined for both HMX [2] and RDX [1] in prior works. If the localization trigger condition is fulfilled, i.e. the von Mises stress becomes larger than a threshold, $\sigma_{vM} > \sigma_{CSSL}$, and the (von Mises) plastic strain increases above a limit, $\varepsilon_{vM} > \varepsilon_{CSSL}$, Eq. (28) is replaced by a flow stress model of the form:

$$\sigma_Y^* = \frac{\sigma_Y + \sigma_{flow}^\infty}{2} + \frac{\sigma_{flow}^\infty - \sigma_Y}{2} \tanh(\varepsilon_{vM}^p) \quad (9)$$

Where $\sigma_Y$ is computed using the unmodified J-C in Eq. (28), $\sigma_{flow}^\infty$ is the value of the flow stress in the amorphized shear band to which the plastically deformed material relaxes as the plastic strain $\varepsilon_{vM}^p$ increases to high values. This model has been calibrated for both RDX and HMX, and further description of the physical basis of this model is detailed in [40]. The expressions for $\sigma_{flow}^\infty$ are provided in Table I.

The above models, along with governing conservation laws (see supplementary information) constitutes a closed system of equations to calculate the thermo-mechanical quantities $\rho, u_i, e, p, S_{ij}$, and $T$ during the elasto-plastic deformation of an isotropic solid material (RDX and HMX) under shock conditions. The system of equations is solved as outlined below.

## 2.3 Numerical algorithms and interfacial conditions

The continuum calculations were performed using the sharp-interface Eulerian multi-material dynamics code SCIMITAR3D [45, 46]. The mass, momentum, and energy conservation equations were cast in the Eulerian framework. The interface between solid energetic material (RDX and HMX) and the pore --treated as vacuum in both the continuum and MD simulations-- was embedded in the fixed, Cartesian grid and is tracked as a sharp interface using the levelset method[47, 48]. The conservation laws for mass, momentum, and energy, along with the evolution of deviatoric stresses, were spatially discretized using a 5$^{th}$-weighted order essentially non-oscillatory (WENO) shock-capturing scheme [49-51]. A 3$^{rd}$-order, total-variation-diminishing Runge-Kutta scheme is used for time-integration [52]. To define and advect material interfaces, a narrow-band levelset-based tracking method was utilized. The use of a levelset function allows for the capability to handle large deformations of interfaces which are inherent to shock-induced pore collapse events. The interfacial conditions between the energetic material and pore surface were modeled by applying the free-surface condition using a modified ghost-fluid method [53, 54]. Detailed descriptions of the numerical algorithms, levelset implementation, and interface treatment are available in previous publications [30, 53-55]. The methods have been validated in previous work [56-59] for a variety of multi-material problems, including pore collapse.

All molecular dynamics simulations were performed using the Large-scale Atomic/Molecular Massively Parallel Simulator (LAMMPS) [60]. A version of the non-reactive, fully flexible forcefield developed by Smith and Bharadwaj [21, 61] was employed for both HMX and RDX, where harmonic functions are used to model covalent bonds, three-center-angles and wag angles. Cut-off distances of 11.0 Å (HMX) and 15.0 Å (RDX) were imposed for all repulsion-dispersion interactions and pairwise Coulombic interactions. The long-range Coulombic interactions were calculated using the particle-particle particle-mesh (PPPM) $k$-space solver described in [62], with a relative error tolerance of $10^{-6}$ (HMX) and $10^{-5}$ (RDX). An additional repulsive function was added to overcome potential simulation instabilities that occur at short interatomic distances due to the diverging potential representing the repulsion-dispersion interactions [63].

## 2.4 Simulation setup and boundary conditions

Figure 1 shows the computational setup employed for the continuum pore collapse simulations. A single circular pore is embedded in a block of either RDX or HMX. The pore diameter ($D_{pore}$) and the corresponding dimensions of the computational domain are varied in the proportions indicated in Figure 1. Pore diameters of 50, 100, 500, and 1000 nm were studied in the continuum simulations, while only 50 nm pores were used in the MD simulations (with the exception of one 100 nm pore calculation for RDX). The block of energetic material impacts a rigid wall at speed $U_p$. Impact velocities of 0.5, 1.0, 1.5, 2.0, 2.5, 3.0 km/s were studied in the continuum calculations for all pore sizes; MD simulations are far more computationally intensive than continuum pore collapse calculations and therefore the entire suite of velocities and pore sizes could not be accessed in the MD dataset. Therefore, with MD, the 50 nm pore was studied at all velocities, but the larger 100 nm pore collapse was only simulated with an impact speed of 1 km/s.

For the simulations employed in this work, the MD samples consist of quasi-2D supercells with a cylindrical pore embedded into the center. While the computational domain and pore arrangement for both materials was nearly the same, some differences in the setup of the HMX and RDX cases were unavoidable due to the difference in their molecular structures. The simulation and model details for 50 nm diameter pore systems are described in detail in prior works HMX [2] and RDX [1]. In addition, one 100 nm pore collapse simulation was performed for RDX only. For this 100 nm diameter pore, the equilibrium dimensions of the RDX unit cell ($a$=1.343 nm, $b$=1.184 nm, $c$=1.077nm) were determined for the Smith-Bharadwaj force field by thermalizing the bulk, periodic crystal to 300 K and 1 atm using NPT simulations. The orthorhombic primary cell was then generated by replication of the unit cell, the final dimensions of the supercell was 462.7 nm × 5.6 nm × 301.6 nm ($337a \times 5b \times 280c$), containing 3,774,400 RDX molecules. Then, atoms were selectively deleted to form a 100 nm cylindrical pore placed directly in the center of the domain. A 10 nm vacuum space was added at the trailing edge of the simulation domain to reduce surface-surface interactions across the periodic boundary. The crystal frames are matched to the Cartesian lab frame such that $a \parallel x$, $b \parallel y$, and $c$ in the $+z$ direction. The supercell was equilibrated in the same manner as the 50 nm RDX system as described in [1] and reverse ballistic impact simulations carried out as in previous works [1, 2].

Simulation setups for both the MD and continuum calculations were nearly identical, but the MD simulations were quasi-2D, while the continuum simulations were performed as fully 2-D by adopting the plane-strain assumption. The MD calculations employ a standard three-dimensional (3D) periodic boundary condition, which stipulates that atomic velocities remain constant for atoms exiting a given face of the primary simulation cell and promptly re-enter the cell at the opposite face. In MD, shocks were generated with a reverse ballistic impact, where a few molecular layers of atoms were frozen along the bottom of the cell to represent a rigid piston. The frozen atoms are held with no velocity or forces throughout the duration of the simulation. Further information regarding the details of the MD shock simulations can be found in prior works [1, 2]. For all continuum simulations, a reflective boundary condition is established at each side of the rectilinear computational domain. The reflective boundary condition on the south, or bottom, edge mimics a rigid wall or a stationary piston in MD. The reflective boundary condition used at the left and right boundaries approximates the periodic boundary condition used in the MD simulations due to the symmetry of the computational domain with respect to the vertical centerline. The lateral boundaries are far enough away from the region of pore collapse and shear band formation during the duration of interest (viz., shock passage across the pore and the resulting collapse) that no wave reflections or other effects of the lateral boundary are detected to affect the dynamics of collapse and hotspot formation.

## 3. RESULTS AND DISCUSSION

Figure 2 shows comparisons between MD and continuum simulations of pore collapse. The pore is 50 nm in diameter and is subject to a moderate strength shock produced by a reverse ballistic impact with particle speed of 1 km/s (Figure 1). Results from two different continuum simulations are depicted, one with a grid resolution of 1 nm and the other with 0.1 nm resolution. Note that in Figure 2 and in other figures to follow, the bulk post-shock temperature is subtracted from the temperature at each spatial location (grid cell center) to emphasize the local increase in temperature due to various mechanisms of energy localization. All three simulations show a lateral pinching-type collapse beginning around ±45° below the horizontal, indicating that the collapse is dominated by plastic deformation for this shock strength or impact velocity. Energy localization due to shear stresses grow from the pore surface in the form of shear bands whose spacings coarsen as time progresses. The shear bands formed in Fig. 2(ii) are more diffuse than in Fig. 2(iii) due to the lower grid resolution but show better agreement with MD. When compared with MD, the "better resolved" 0.1 nm resolution in Fig. 2(iii) produces a large number of sharper shear bands at higher temperatures. This can also be seen in the histogram comparing the temperature distributions of the three calculations in Fig. 2 (iv). The 0.1 nm resolution temperature field shows discrepancies from MD in the range of 100-300K. At the instant of collapse, the continuum simulation using 1 nm grid resolution aligns much better with MD both in shear band characteristics and maximum temperature of the hotspot. Additionally, the rate of pore collapse in Fig. 2(v) is in better alignment with MD when the coarser grid resolution of 1 nm is employed. Note that the 0.1 nm grid spacing produces sharper hot spot features, including shear bands since the grid is "super-resolved" in this case, i.e., finer than the smallest length scales relevant to the physics of the problem, viz., the molecular size. Therefore, the shear bands obtained at this higher resolution are unphysical, leading to the lack of agreement with the MD result. The results to follow reinforce this viewpoint and support the use of a 1 nm grid resolution for both materials and across the range of pore sizes and impact velocities covered in this work.

An identical, reverse ballistic pore collapse setup was used in Figure 3 to compare hotspot characteristics in HMX. As seen in RDX, a weak shock with particle speed of 1 km/s causes the pore to collapse in a visco-plastic manner in HMX. Similar to the RDX case, the shear band structure for 1 nm grid resolution (Fig. 3(ii)) is more diffuse in nature than for the 0.1 nm resolution in Fig. 3(iii); in the latter case the number and temperature of shear bands is higher. Also, in Fig. 3(ii), the maximum temperature in the core of the hotspot corresponds better with what is seen in MD for the 1 nm resolution than for the 0.1 nm resolution. These observations are supported by comparing the temperature distributions for the three calculations shown in Fig. 3(iv). Again, the 0.1 nm grid resolution produces shear band temperatures that are hotter than MD which explains the difference in the distributions in the lower temperature range. The 1 nm grid resolution brings the hot spot temperatures into better agreement with MD primarily due to the more diffuse nature of the shear localization. For HMX, unlike RDX, the collapse rate does not seem to be strongly dependent on the grid resolution as both continuum calculations show pore collapse at a rate slower than MD, as seen in Fig. 3(v).

Figure 4 compares temperature contours and distributions for a larger void in RDX, of 100 nm diameter, and for pore collapse at an impact speed of 1 km/s. Like the 50 nm pore in Figure 2, the larger 100 nm pore collapses in a visco-plastic manner. The MD calculation, Fig. 4(i), preserves more of the bottom surface of the pore due to weaker material jetting, while both continuum simulations show stronger jetting leading to slightly different pore shapes at the instant shown in the middle column, corresponding to $A/A_0$ = 25%. The resulting shear band network generated in MD again shows diffuse bands where material is heated to temperatures that exceed the surrounding post-shock bulk temperature by ~300K. A similar pattern and temperature are seen in the continuum simulation using 1 nm grid resolution in Fig. 4(ii). However, when using 0.1 nm resolution in Fig. 4(iii), the shear bands are much hotter, in the range of ~600 K above the bulk temperature, with certain areas in the shear band rising to 800 – 900 K. Additionally, for the more refined grid, intense shear banding can be seen in the bulk material away from the pore. Shear bands in the bulk can also be seen to occur in the MD simulations; however, the bands are more abundant, sharper and hotter for the 0.1 nm resolution. The maximum temperature of the hot spot, depicted by the red region in the contours is in good qualitative agreement, in terms of the magnitude and distribution, between MD (Fig. 4(i)) and the 1 nm resolution continuum calculation (Fig. 4(ii)); whereas the 0.1 nm resolution (Fig. 4(iii)) case shows more of an arc-like shape that holds the hottest temperature. The high temperature in the core of the hotspot is an important characteristic as these high temperature areas approaching 1000K are the most likely to initiate chemical reactions in the material. The amplified energy localization and temperature for the refined grid (super-resolved, 0.1 nm) case is reflected in the temperature distributions in Fig. 4(v). The super-resolved case leads to higher temperature across the entire distribution when compared to MD, due to the intense shear bands as well as the larger area of the collapse generated hotspot. The 1 nm resolution case in contrast is in better agreement with the MD temperature distribution even in the tail of the distribution. However, counterintuitively, the continuum pore collapse process in time is, for both levels of grid refinement, in good agreement with each other but both are different from the MD case. Thus, the agreement in terms of the shear band patterns and the hotspot temperatures fields are not necessarily in alignment with the details of spatial or temporal pore collapse profiles.

Pore size and impact speed have been shown to play a significant role in collapse behavior and the resulting hot spot characteristics [64-66]. To study these effects, continuum simulations were conducted for varying pore diameters and particle velocities while fixing the grid resolution at the molecular length scale, i.e., at 1 nm grid size for all cases. Four different pore sizes (50 nm, 100 nm, 500 nm, and 1000 nm) and six different particle velocities (0.5 km/s, 1 km/s, 1.5 km/s, 2 km/s, 2.5 km/s, and 3 km/s) were chosen to investigate the full range of shock strengths. Figures 5 and 6 show collapse events with an impact speed of 0.5 km/s for the full range of pore sizes in RDX and HMX, respectively. In both RDX and HMX, regardless of pore size, a sideways pinching collapse is seen for this low velocity case. Unlike in the classical view of hydrodynamic jetting in void collapse, material jets occur laterally and collide in the center to produce the maximum temperature in the hot spot. At this low impact speed, the collapse is viscoplastic in nature and the hotspot is dominated by the temperature rise due to plastic deformation localized in the shear bands. I.e., in contrast to previous calculations of pore collapse and hotspot formation the current atomistics-consistent models produce hotspots that occupy a larger region around the pore and also contain a significant fraction of energy in the shear band as opposed to the region heated by the collapse of the pore. In recent work [1, 2] we showed that for lower velocities energy localization in the shear band predominates over the localization in the tail of the hotspot temperature distribution, i.e., in the collapse zone. As pore size increases, the number and sharpness of shear bands also increases. By maintaining the grid resolution at 1 nm, the shear bands become finer as the domain size continues to increase, as seen by comparing the cases from top to bottom in each column in Figures 5 and 6. By comparing the two figures it can also be seen that the differences between RDX and HMX are small in terms of the pore shape as well as shear band characteristics at the three instants shown, across all the pore sizes and velocities studied in this work.

Figures 7 and 8 show collapse for the same range of pore sizes but for a higher impact speed of 1.0 km/s. Fig. 7, displaying the RDX simulations, also includes an MD calculation of a 100 nm pore (Fig. 7(ii)). For this moderate shock strength, the collapse occurs not entirely in a sideways pinching mode but rather via a diagonal, dual-jetting action beginning at locations at the bottom of the collapsing pore. This is reflected in MD as well as the continuum simulations. The overall shear banding behavior seen in MD is also well captured by the continuum simulations. The shear bands are diffuse in nature for the smaller void sizes and become more refined as the pore size increases; note that MD data is unavailable at this time for the larger void sizes since such calculations are computationally intensive. The temperature and the spacing between the bands are also consistent between the MD and continuum calculations (compare Fig. 7(i) with Fig. 7(iii), and Fig. 7(ii) with Fig. 7(iv)). However, the MD cases show weak shear localization occurring in the bulk of the material away from the pore (see Figs. 7(i) and (ii)); these bands are more prominent in the larger 100 nm pore collapse case (Fig. 7(ii)). The continuum simulations fail to capture these bulk reticulated structures as the shear bands arise primarily from the surface of the pore in the continuum model. This is likely due to the smaller stiffness of the M-JC model used in the continuum calculations, as evident from the pore collapse versus time plots displayed for the cases in this paper; modifications to the localization criterion in the model may be needed to remedy this discrepancy away from the pore. Similar to the low velocity case, the pore collapse and shear band behaviors for HMX, in Fig. 8, show that the differences between the two materials is rather modest.

Pore collapse calculations at an impact speed of 2.0 km/s are shown in Figures 9 and 10, for RDX and HMX respectively. For this higher strength shock, the pore collapse is expected to be pressure-driven and approaching the hydrodynamic regime. This behavior is shown in the MD case for both RDX (Fig. 9) and HMX (Fig. 10). There is still, however, some effect of localization, which manifests as a softening and double jet formation at the lower surface of the pore, showing a fully hydrodynamic state has not yet been reached. For the continuum calculations, the double involution and localization seen at the lower surface of the pore is more pronounced than in the MD case. As the pore size increases, many shear bands begin to appear in both RDX and HMX. The structure of the shear bands in the two materials is different, however, with the RDX case showing a reticulated structure as observed in the MD case shown in Fig. 7(ii). The HMX shear bands do not display such reticulation. These differences in the shear band patterns away from the pore collapse region are interesting and further examination of the localization models in the two materials is warranted to examine the physics underlying such patterns. In both RDX and HMX the resulting hotspot temperature in the continuum calculations is hotter than what is seen in MD (compare Fig. 9(i) with Fig. 9(ii), and Fig. 10(i) with Fig. 10(ii)). This shows the current strength and localization models may need further modification to better capture pore collapse and hot spots associated with strong shocks. The models also need to be further tested for large pore sizes and range of shock strengths. Unfortunately, MD calculations of large pores, in excess of 100 nm diameter are extremely computationally intensive. For example, the computational times for continuum calculations of 50nm pores are in the 300 CPU hours range whereas MD simulations require approximately 1.4 million CPU hours. In contrast, while 500nm pores can be simulated in the continuum case in 10,000 CPU hours, MD simulations take roughly 60 million CPU hours to complete. Nevertheless, despite the rather heavy computational load for larger pore sizes, effort is ongoing to perform MD calculations on 250nm and 500nm pores and will be reported on in a future publication.

To perform a quantitative assessment, various hot spot and pore collapse metrics have been used to compare the continuum models with MD. Using all of the MD data for RDX (50 nm diameter pore at six impact speeds, and 100 nm diameter pore at 1 km/s) and HMX (50 nm diameter pore at three impact speeds), quantitative comparisons have been made with the entire range of continuum pore collapse calculations. Firstly, with regard to the plots shown in Figure 11, row(a) depicts comparison of the pore collapse duration in the two materials, with RDX in the left column and HMX on the right. The x-axis covers the range of impact velocities and the y-axis is a non-dimensional collapse time. By non-dimensionalizing the collapse time, a unified comparison can be made independent of pore size. The collapse duration, $\tau_{collapse}$, is the time interval starting from when the shock first reaches the pore to the point of complete collapse, i.e., when the pore area reaches zero. To non-dimensionalize, the collapse duration is divided by a shock passage time scale $\tau^*$ determined by:

$$\tau^* = \frac{D_{pore}}{u_s} \qquad (10)$$

Here $D_{pore}$ is the pore diameter and $u_s$ is the shock speed.

Several observations can be made from Figure 11(a) on the collapse time of pores in the two materials. For both RDX and HMX, the collapse times from continuum and MD calculations are in closer agreement at higher impact speeds of 1 km/s. The non-dimensional collapse rate increases as the pore diameter increases and appears to saturate for larger pore size. The non-dimensional collapse rate also saturates as the impact velocity (shock strength) increases. This saturation behavior is due to the increasingly hydrodynamic nature of the collapse as pore size or shock strength are increased. For weaker shocks ($u_p$=0.5 km/s) the MD pores collapse at a slower rate than in the continuum calculation. The discrepancy between MD and continuum collapse times for the 0.5 km/s impact is larger for RDX than HMX. This discrepancy at lower velocities is likely because the M-JC model leads to a material with lower stiffness than the atomistic case.

The next row of plots in Fig. 11(b) compares the spatially averaged hotspot temperature for the range of simulations presented in this work. The hotspot temperature was computed in the following way:

$$T_{hs,avg} = \frac{\int_{\Omega_{hs}} T_{hs} d\Omega_{hs}}{A_{hs}} \qquad (11)$$

Here, $T_{hs}$ and $A_{hs}$ refer to the temperature field in the hotspot and the area occupied by the hotspot, respectively. Temperatures are averaged over the area $\Omega_{hs}$, which is defined as the hot spot region, which spans the regions in the domain where the temperature exceeds the post-shock bulk temperature, $T_{bulk}$. As seen in Fig. 11(b), for HMX, the continuum model does well in capturing MD-consistent hot spot temperatures over the range of impact speeds where MD data is available. The full range of continuum pore sizes produces hotspot temperatures within a few hundred Kelvin of the ground truth. The discrepancies between continuum and MD hotspot temperatures, like the collapse times above, are higher for RDX as seen in Figure 11(b) (left) especially for the smaller (50 nm and 100 nm) pores at higher velocities. It is interesting to note that, contrary to the trend in the pore collapse time, the hotspot temperature for the low velocities are in good agreement with MD, for RDX as well as HMX. Also, the larger 500 nm and 1000 nm pores in RDX yield averaged hotspot temperatures in close agreement with MD; note that the MD calculation was performed only for the 50 nm pore. The hotspot temperatures also appear to saturate in value as the pore size and impact velocities increase, indicative of a predominantly hydrodynamic mode of collapse.

From the two metrics of pore collapse and hotspot formation above, namely pore collapse time (Figure 11 (a) and 11(b)), the primary discrepancies with the ground truth are at small pore diameters and high shock strengths (impact velocities). It has already been noted that the continuum yield stress model may need to be calibrated more closely for stronger shocks to achieve even closer agreement with MD. In the current model, the Johnson-Cook model results in a lack of material softening in the region away from the pore. Such softening at high shock strengths has been noted in a recent work [67] due to the nucleation of multitudes of nano-scale shear bands. This may lead to less intense energy localization and therefore lower hotspot temperatures at higher shock strengths, potentially correcting the discrepancy for RDX in Figure 11(b) (left). The softening away from the pore could also lead to accelerated collapse of the pore in Figure 11(a) bringing the continuum and MD results in closer agreement for that metric as well. These issues are being examined in ongoing work.

Figure 11(c) shows the variation of the ratio of energy contained in the shear bands to the total hotspot energy plotted against the impact speed, providing insights on energy localization modalities during pore collapse. First, the total energy contained in the hot spot is given by:

$$e_{hs} = c_v * \int_{\Omega_{hs}} \rho_{hs} T_{hs} d\Omega_{hs} \qquad (12)$$

$\rho_{hs}$ and $T_{hs}$ are the density and temperature within the hot spot and the integral is over the hot spot region $\Omega_{hs}$. To align with MD, the specific heat $c_v$ is constant. Next, the energy contained in the shear bands $e_{sb}$ is computed in a similar manner as above except that the internal energy is integrated over the region $\Omega_{sb}$ occupied by the shear band only. The shear band region is chosen by analyzing the temperature distributions for each case from the histograms shown in Figures 2, 3, and 4. Two local maxima are identified for each distribution, the first accounting for temperatures due to localization in shear bands and the second at higher temperatures attributed to the conversion of kinetic energy to heat which is due to the material jet collision. The final hotspots in Figures 2,3, and 4 also provide visual representations of these two maxima in the temperature distributions. The shear bands contain material at temperatures in the range of 100K-700K above $T_{bulk}$ while the collapse zones are in the range of 700K and above. In this work, the shear band region includes temperatures just above the bulk temperature until the end of the first peak (~500-700K above $T_{bulk}$ depending on the impact velocity). The ratio of energies $e_{sb}/e_{hs}$ is plotted for each impact velocity and pore diameter in Figure 11 (c). For both RDX and HMX, at the lower impact velocities, the hotspot energy is predominantly localized in shear bands. This is expected for weaker shocks where the collapse is primarily driven by plastic deformation. However, even at the highest velocity of 3 km/s the shear bands contain a larger proportion of the energy than the collapse zone. It is interesting to note that the distribution of energy between the shear bands and the collapse zone are well captured for small voids at all velocities, as evident from the agreement with MD across the velocity range. There is a significant deviation from the MD trend for the larger pores and at higher velocities for RDX. From Figure 9, this discrepancy may be attributed to the occurrence of the network of reticulated shear bands in the material away from the pore collapse site for the larger pores (see Figures 9(iv) and 9(v)). In contrast, the model does not yield a network of crisscrossed bands for HMX (see Figure 10(iv) and 10(v)). In the absence of MD data for large pores, approaching 1 micron in size, the validity of these shear band structures remains to be established although some MD simulations have shown the existence of such reticulated structures [67, 68]; such large pore calculations are underway in ongoing work.

Finally, we assess the implications of the above observations on the energy distribution in shear bands versus the collapse zone for hotspot ignition and growth. Note that the MD as well as continuum calculations in this paper are performed for an inert material. However, reactions are initiated in RDX as well as HMX at temperatures of 1000K or higher, i.e., although the shear bands localize a significant portion of the energy left behind by the shock, reactions may not be triggered in the entire hotspot. Only the tail of the temperature distributions in the hotspot, i.e., the core of the hotspots in the final collapse zone may lead to ignition. To this end, the plots in Figure 11(d) show a predicted reaction time scale as a function of the impact velocity. The reaction time scale is calculated using single-step Tarver reaction chemistry model reaction parameters for RDX and HMX decomposition [28, 29]. For each case of pore size and impact velocity, an effective reaction rate was computed as an averaged quantity over the hotspot area as follows:

$$\overline{Q_r} = \frac{\int_{\Omega_{hs}} Z e^{-\frac{E_a}{RT_{hs}}} d\Omega_{hs}}{A_{hs}} \quad (13)$$

Where $E_a$ is the activation energy and $Z$ is the collision frequency with values for RDX and HMX given in Table II. Note that the Arrhenius form of the reaction rate is effectively a low-pass filter for the temperature field, in that the contributions to the integral in the numerator is weighted towards the high temperature regions in the hotspot. The effective time scale of reaction can then be computed from taking the inverse of the averaged reaction rate $\overline{Q_r}(s^{-1})$.

$$t_{reaction} = \frac{1}{\overline{Q_r}} \quad (14)$$

Figure 11(d) shows a strong variation in the calculated effective reaction timescale for both RDX (left) and HMX (right). The reaction time scale for the weaker shock (e.g., 0.5 km/s impact velocity) is rather large and is unlikely to yield a critical hotspot [25]. As the shock strength increases the reaction time scales decreases precipitously and critical conditions leading to ignition and growth become possible [69]. For larger pore sizes the reaction time scales decrease significantly compared to the small pore sizes due to the stronger and extensive localization in the shear bands for the larger pores. However, even for the large pores the impact velocities have to be sufficiently high to achieve reaction time scales small enough to initiate critical hotspots. Therefore, while shear bands may contain the bulk of the energy localization in the hotspot formed around the collapsed pore, their potential to trigger reactions is low, due to the low temperatures in the bands. However, the shear bands may offer favorable pathways for reaction propagation once ignition occurs in the high temperature collapse zone, thus accelerating the growth of the hotspot. These effects will need to be investigated by performing reactive pore collapse calculations using the current model, which is being pursued in ongoing work.

## 4. CONCLUSIONS

Hotspot formation through energy localization at defects plays a key role in the shock sensitivity of energetic materials such as HMX and RDX. In this work, MD-informed material models for HMX and RDX were employed in continuum calculations and compared to corresponding MD data on hotspot temperatures, shear localization and pore collapse profiles (where such calculations were feasible).

Continuum pore collapse simulations were performed in both materials for a range of pore sizes and impact speeds. An unresolved issue in continuum simulations of pore collapse has been the appropriate grid resolution for obtaining physically correct collapse and energy localization. Conventional computational mechanics guidelines advise use of grid refinement to levels that provide grid independent solutions. However, using such criteria led to unphysical shear bands and discrepancy with the ground truth MD calculations. Through comparisons with tandem MD calculations, it was shown that adopting a cutoff (molecular) length scale of 1 nm for continuum grid resolution produces better agreement with MD temperatures fields and shear localization across the pore size and shock strength range studied in this work. Grids finer than 1 nm produce unphysical features that can significantly impact collapse dynamics and hot spot temperatures. Employing a physics-guided mesh resolution, cutoff at molecular length scale, leads to shear bands that are diffuse in the case of small pores whereas for larger pores fine, nano-scale shear bands are obtained; in the case of RDX these shear bands are seen to adopt a reticulated pattern which does not occur in the case of HMX. The physical correctness of these finer, dense shear band patterns seen for the larger pore sizes need to be confirmed by performing MD simulations of large pores approaching the micron size range.

Simulations over a range of shock strengths (i.e., reverse ballistic impact velocities) show that for weak shocks, the localization of plastic flow in shear bands is the main driver of pore collapse. For both HMX and RDX, shocks with impact speeds of 1.5 km/s or less show at least 90% or more of the total hot spot energy being contained in shear bands. This indicates the predominance of viscoplastic collapse where the energy is largely localized in shear bands. The highest discrepancy between the continuum and MD calculations is observed in the strong shock regimes. At high pressures, the strain-rate dependency in the Johnson-Cook strength model can lead to the bulk material, away from the pore, becoming too stiff. This will lead to enhanced shear band formation and therefore to a more plasticity-dominated collapse, which deviates from the expected hydrodynamic collapse for stronger shocks seen in MD. The discrepancies are also seen in the collapse of larger pores, where the continuum simulations produce shear bands in the bulk of the material which is not seen in MD simulations. These remaining discrepancies suggest that the atomistics-informed modified Johnson-Cook (M-JC) strength model used in the continuum simulations needs to be more finely calibrated over a wider range of length scales and shock strengths.

Ongoing work seeks to address the deficiencies noted above with regard to the high shock strength response of RDX and HMX. MD simulations for large pore sizes are computationally intensive; therefore, the large pore simulations reported in this work await validation against MD simulations which are ongoing. In addition, as a starting point, this work limited itself to the simulation of the inert collapse of pores in RDX as well as HMX, both to develop the material models for the two materials from atomistic calculations and to examine the issue of the resolution to be employed in the continuum calculations. Reactive pore collapse simulations are being performed to study the effect of shear localization, pore size and shock strength effects using the present atomistics-consistent models and will be reported in a forthcoming publication.

**ACKNOWLEDGMENTS**

This work was done as part of the doctoral dissertation of JH, who was sponsored by the Army Research Laboratory and was accomplished under Cooperative Agreement Number W911NF-19-2-0110 and an ARO grant number W911NF-22-2-0164. CO's work was supported by a DoD/AFOSR MURI grant: federal award number FA9550-19-1-0318, program managers Dr. Martin Schmidt (2019-2022) and Drs. Chiping Li and Derek Barbee (2022-present). TS gratefully acknowledges an AFOSR DURIP equipment grant for local HPC resources at the University of Missouri: federal award number FA9550-20-1-0205. The views and conclusions contained in this document are those of the authors and should not be interpreted as representing the official policies, either expressed or implied, of the Army Research Laboratory or the U.S. Government. The U.S. Government is authorized to reproduce and distribute reprints for Government purposes notwithstanding any copyright notation herein.

**DATA AVAILABILITY**

The data that support the findings of this study are available from the corresponding author upon reasonable request.

**5. REFERENCES**

1. Herrin, J., et al., *Pore collapse, shear bands, and hotspots using atomistics-consistent continuum models for RDX (1,3,5-trinitro-1,3,5-triazinane): Comparison with molecular dynamics calculations.* Journal of Applied Physics, 2024. **136**(13).
2. Nguyen, Y.T., et al., *Continuum models for meso-scale simulations of HMX (1,3,5,7-tetranitro-1,3,5,7-tetrazocane) guided by molecular dynamics: Pore collapse, shear bands, and hotspot temperature.* Journal of Applied Physics, 2024. **136**(11).
3. Barnes, B.C., Brennan, J. K., Byrd, E. F., Izvekov, S., Larentzos, J. P., & Rice, B. M. , *Toward a Predictive Hierarchical Multiscale Modeling Approach for Energetic Materials*, in *Computational Approaches for Chemistry Under Extreme Conditions*, N. Goldman, Editor. 2019, Springer International Publishing: Cham. p. 229-282.
4. Barnes, B.C., Leiter, K. W., Larentzos, J. P., Brennan, J. K., *Forging of hierarchical multiscale capabilities for simulation of energetic materials.* Propellants Explosives Pyrotechnics, 2020. **45**: p. 177-195.
5. Barnes, B.C., Spear, C. E., Leiter, K. W., Becker, R., Knap, J., Lísal, M., & Brennan, J. K. . *Hierarchical multiscale framework for materials modeling: Equation of state implementation and application to a Taylor anvil impact test of RDX.* in *AIP Conference Proceedings*. 2017. American Institute of Physics.
6. Rai, N.K., Sen, O.,Udaykumar, H. S., *Macro-scale sensitivity through meso-scale hotspot dynamics in porous energetic materials: Comparing the shock response of 1,3,5-triamino-*

*2,4,6-trinitrobenzene (TATB) and 1,3,5,7-tetranitro-1,3,5,7-tetrazoctane (HMX).* Journal of Applied Physics, 2020. **128**(8).
7. Udaykumar, H.S., et al., *Unified Approach for Meso-Informed Burn Models in Shocked Energetic Materials.* Journal of Propulsion and Power, 2020. **36**(5): p. 655-667.
8. Udaykumar, H., Y.T. Nguyen, and P. Kumar Seshadri, *Physically Evocative Meso-Informed Burn Model: Topology of Evolving Hotspot Fields.* Journal of propulsion and power, 2022. **38**(6): p. 920-934.
9. Kumar Rai, N. and H.S. Udaykumar, *A Eulerian level set‐based framework for reactive meso‐scale analysis of heterogeneous energetic materials*, in *Dynamic Damage and Fragmentation*. 2018, Wiley. p. 387-416.
10. Rai, N.K., *Numerical framework for mesoscale simulation of heterogeneous energetic materials*. 2015: The University of Iowa.
11. Johnson, B.P., et al., *Observing Hot Spot Formation in Individual Explosive Crystals Under Shock Compression.* J Phys Chem A, 2020. **124**(23): p. 4646-4653.
12. Olokun, A., et al., *Examination of Local Microscale-Microsecond Temperature Rise in HMX-HTPB Energetic Material Under Impact Loading.* Jom, 2019. **71**(10): p. 3531-3535.
13. Bassett, W.P. and D.D. Dlott, *High dynamic range emission measurements of shocked energetic materials: Octahydro-1,3,5,7-tetranitro-1,3,5,7-tetrazocine (HMX).* J. Appl. Phys., 2016. **119**: p. 225103.
14. Curtis, A.D., et al., *Laser-driven flyer plates for shock compression science: launch and target impact probed by photon Doppler velocimetry.* Rev Sci Instrum, 2014. **85**(4): p. 043908.
15. Bassett, W.P., et al., *Shock Initiation Microscopy with High Time and Space Resolution.* Propellants, Explosives, Pyrotechnics, 2020. **45**: p. 223-235.
16. Dlott, D.D., et al., *Shock Compression Microscopy: Shocked Materials with High Time and Space Resolution*, in *LLNL-PROC-782266*. 2019.
17. Bassett, W.P. and D.D. Dlott, *Multichannel emission spectrometer for high dynamic range optical pyrometry of shock-driven materials.* Rev Sci Instrum, 2016. **87**(10): p. 103107.
18. Hamilton, B.W., et al., *Sensitivity of the shock initiation threshold of 1, 3, 5-triamino-2, 4, 6-trinitrobenzene (TATB) to nuclear quantum effects.* The Journal of Physical Chemistry C, 2019. **123**(36): p. 21969-21981.
19. Rice, B.M., et al., *Parameterizing complex reactive force fields using multiple objective evolutionary strategies (MOES): part 2: transferability of ReaxFF models to C–H–N–O energetic materials.* Journal of chemical theory and computation, 2015. **11**(2): p. 392-405.
20. Zybin, S.V., et al. *ReaxFF reactive molecular dynamics: Coupling mechanical impact to chemical initiation in energetic materials.* in *2010 DoD High Performance Computing Modernization Program Users Group Conference.* 2010. IEEE.
21. Smith G. D. and Bharadwaj R. K., *Quantum chemistry based force field for simulations of HMX.* The Journal of Physical Chemistry B, 1999. **103**(18): p. 3570-3575.
22. Sen, O., Gaul, N. J., Choi, K. K., Jacobs, G., Udaykumar, H. S., *Evaluation of kriging based surrogate models constructed from mesoscale computations of shock interaction with particles.* Journal of Computational Physics, 2017. **336**: p. 235-260.
23. Sen, O., et al., *Multi-scale shock-to-detonation simulation of pressed energetic material: A meso-informed ignition and growth model.* Journal of Applied Physics, 2018. **124**(8): p. 085110.
24. Rai, N.K. and H.S. Udaykumar, *Void collapse generated meso-scale energy localization in shocked energetic materials: Non-dimensional parameters, regimes, and criticality of hotspots.* Physics of Fluids, 2019. **31**(1): p. 016103.
25. Tarver Craig M., Chidester Steven K., and Nichols Albert L., *Critical conditions for impact- and shock-induced hot spots in solid explosives.* The Journal of Physical Chemistry, 1996. **100**(14): p. 5794-5799.
26. Rai, N., I. Schweigert, and H. Udaykumar. *A comparative study of chemical kinetics models for HMX in mesoscale simulations of shock initiation due to void collapse.* in *APS Shock Compression of Condensed Matter Meeting Abstracts.* 2015.

27. Parepalli, P., et al., *Multi scale models for heterogenous energetic materials III, "Multi-scale modeling of shock initiation of a pressed energetic material III: Effect of Arrhenius chemical kinetic rates on macro scale shock sensitivity.* Journal of Applied Physics, 2024. **138**(8).
28. Tarver C. M. and Tran T. D., *Thermal decomposition models for HMX-based plastic bonded explosives.* Combustion and Flame, 2004. **137**(1-2): p. 50-62.
29. McGuire R.R. and Tarver C. M. *Chemical decomposition models for the thermal explosion of confined HMX, TATB, RDX, and TNT explosives*. in *Seventh Symposium (International) on Detonation*. 1981.
30. Kapahi A., Sambasivan S., and Udaykumar H. S., *A three-dimensional sharp interface Cartesian grid method for solving high speed multi-material impact, penetration and fragmentation problems.* Journal of Computational Physics, 2013. **241**: p. 308-332.
31. Kapahi A. , et al., *Parallel, sharp interface Eulerian approach to high-speed multi-material flows.* Computers & Fluids, 2013. **83**: p. 144-156.
32. Sambasivan, S.K., *A sharp interface Cartesian grid hydrocode*, in *Mechanical Engineering*. 2010, University of Iowa.
33. Sambasivan, S.K.U.H.S., *Sharp interface simulations with Local Mesh Refinement for multi-material dynamics in strongly shocked flows.* Computers & Fluids, 2010. **39**(9): p. 1456-1479.
34. Tran L. and Udaykumar H. S., *Simulation of void collapse in an energetic material, Part II: Reactive case.* Journal of Propulsion and Power, 2006. **22**(5): p. 959-974.
35. Tran L. and Udaykumar H. S., *Simulation of void collapse in an energetic material, Part I: Inert case.* Journal of Propulsion and Power, 2006. **22**(5): p. 947-958.
36. Kroonblawd M. P and Austin R. A., *Sensitivity of pore collapse heating to the melting temperature and shear viscosity of HMX.* Mechanics of Materials, 2021. **152**: p. 103644.
37. Kroonblawd, M.P. and H.K. Springer, *Predicted Melt Curve and Liquid Shear Viscosity of RDX up to 30 GPa.* Propellants, Explosives, Pyrotechnics, 2022. **47**(8): p. e202100328.
38. Kechin, V., *Thermodynamically based melting-curve equation.* Journal of Physics: Condensed Matter, 1995. **7**(3): p. 531.
39. Pereverzev A. and Sewell T. D., *Elastic coefficients of β-HMX as functions of pressure and temperature from Molecular Dynamics.* Crystals, 2020. **10**(12): p. 1123.
40. Sen, O., et al., *An Eulerian crystal plasticity framework for modeling large anisotropic deformations in energetic materials under shocks.* . Journal of Applied Physics, 2022. **132**(18): p. 185902.
41. Sen, O., et al., *Johnson–Cook yield functions for cyclotetramethylene-tetranitramine (HMX) and cyclotrimethylene-trinitramine (RDX) derived from single crystal plasticity models.* Journal of Applied Physics, 2024. **135**(14): p. 145901.
42. Khan Mohammad and Picu Catalin R., *Shear localization in molecular crystal cyclotetramethylene-tetranitramine (β-HMX): Constitutive behavior of the shear band.* Journal of Applied Physics, 2020. **128**(10): p. 105902.
43. Wood M. A., et al., *Multiscale modeling of shock wave localization in porous energetic material.* Physical Review B, 2018. **97**(1): p. 014109.
44. Z.C. Zhang, M. Khan, and C.R. Picu, *Mechanism-informed constitutive modeling of molecular crystal cyclotetramethylene tetranitramine (β-HMX).* International Journal of Plasticity, 2023. **169**: p. 103722.
45. Rai, N.K., M.J. Schmidt, and H.S. Udaykumar, *High-resolution simulations of cylindrical void collapse in energetic materials: Effect of primary and secondary collapse on initiation thresholds.* Physical Review Fluids, 2017. **2**(4): p. 043202.
46. Rai N. K. and Udaykumar H. S., *Three-dimensional simulations of void collapse in energetic materials.* Physical Review Fluids, 2018. **3**(3): p. 033201.
47. Albert, S.J., *Level set methods and fast marching methods: evolving interfaces in computational geometry, fluid mechanics, computer vision, and materials science*. Vol. 3. 1999: Cambridge university press.
48. Sethian J. A. and S. Peter, *Level Set Methods for Fluid Interfaces.* Annual Review of Fluid Mechanics, 2003. **35**(1): p. 341-372.
49. Shu C. W. and Osher S., *Efficient implementation of essentially non-oscillatory shock-capturing schemes.* Journal of computational physics, 1988. **77**(2): p. 439-471.

50. Shu C. W., et al., *High-Order Eno Schemes Applied to 2-Dimensional and 3-Dimensional Compressible Flow.* Applied Numerical Mathematics, 1992. **9**(1): p. 45-71.
51. Shu Chi-Wang and Osher Stanley, *Efficient implementation of essentially non-oscillatory shock-capturing schemes, II.* Journal of Computational Physics, 1989. **83**(1): p. 32-78.
52. Gottlieb S. and Shu C.W., *Total variation diminishing Runge-Kutta schemes.* Mathematics of computation, 1998. **67**(221).
53. Sambasivan S. and Udaykumar H. S., *An evaluation of ghost fluid methods for strong shock interactions I: Fluid-fluid interfaces.* AIAA Journal, 2009. **47**(12): p. 2907-2922.
54. Sambasivan S. K. and Udaykumar H. S., *Ghost fluid method for strong shock interactions Part 2: Immersed solid boundaries.* Aiaa Journal, 2009. **47**(12): p. 2923-2937.
55. Udaykumar, H., Tran, L., Shyy, W., Vanden, K., Belk, D. *A combined immersed interface and ENO shock capturing method for multimaterial impact dynamics.* in *Fluids 2000 Conference and Exhibit.* 2000.
56. Udaykumar, H.S., S. Krishnan, and S. Marella, *Adaptively refined, parallelised sharp interface Cartesian grid method for three-dimensional moving boundary problems.* International Journal of Computational Fluid Dynamics, 2009. **23**(1): p. 1-24.
57. Seshadri, V., R. Mittal, and H.S. Udaykumar. *Vortex induced auto-rotation of a hinged plate: a computational study.* in *ASME/JSME 2003 4th Joint Fluids Summer Engineering Conference.* 2003. American Society of Mechanical Engineers Digital Collection.
58. Udaykumar, H.S., R. Mittal, and P. Rampunggoon, *Interface tracking finite volume method for complex solid-fluid interactions on fixed meshes.* Communications in Numerical Methods in Engineering, 2002. **18**(2): p. 89-97.
59. Das, P., et al., *A sharp interface Cartesian grid method for viscous simulation of shocked particle-laden flows.* International Journal of Computational Fluid Dynamics, 2017. **31**(6-8): p. 269-291.
60. Plimpton Steve, Crozier Paul, and Thompson Aidan, *LAMMPS-large-scale atomic/molecular massively parallel simulator.* Sandia National Laboratories, 2007. **18**: p. 43.
61. Bedrov, D., et al., *Molecular dynamics simulations of HMX crystal polymorphs using a flexible molecule force field.* Journal of Computer-Aided Materials Design, 2002. **8**(2-3): p. 77-85.
62. Hockney Roger W. and James W. Eastwood, *Computer simulation using particles.* 2021, Boca Raton: crc Press.
63. Li C. Y., Hamilton B. W., and S. A., *Hotspot formation due to shock-induced pore collapse in 1,3,5,7-tetranitro-1,3,5,7-tetrazoctane (HMX): Role of pore shape and shock strength in collapse mechanism and temperature.* Journal of Applied Physics, 2020. **127**(17).
64. Roy S., et al., *Modeling mesoscale energy localization in shocked HMX, Part II: training machine-learned surrogate models for void shape and void-void interaction effects.* Shock Waves, 2020. **30**(4): p. 349-371.
65. Nguyen, Y., et al., *Multi-scale modeling of shock initiation of a pressed energetic material I: The effect of void shapes on energy localization.* Journal of Applied Physics, 2022. **131**(5).
66. Nassar, A., et al., *Modeling mesoscale energy localization in shocked HMX, part I: machine-learned surrogate models for the effects of loading and void sizes.* Shock Waves, 2019. **29**: p. 537-558.
67. Hamilton, B.W. and T.C. Germann, *High pressure suppression of plasticity due to an overabundance of shear embryo formation.* npj Computational Materials, 2024. **10**(1): p. 147.
68. Eason, R.M. and T.D. Sewell, *Molecular Dynamics simulations of the collapse of a cylindrical pore in the energetic material α-RDX.* Journal of Dynamic Behavior of Materials, 2015. **1**: p. 423-438.
69. Nguyen, Y.T., P.K. Seshadri, and H. Udaykumar, *Physically evocative meso-informed sub-grid source term for energy localization in shocked heterogeneous energetic materials.* Journal of Applied Physics, 2023. **134**(16).

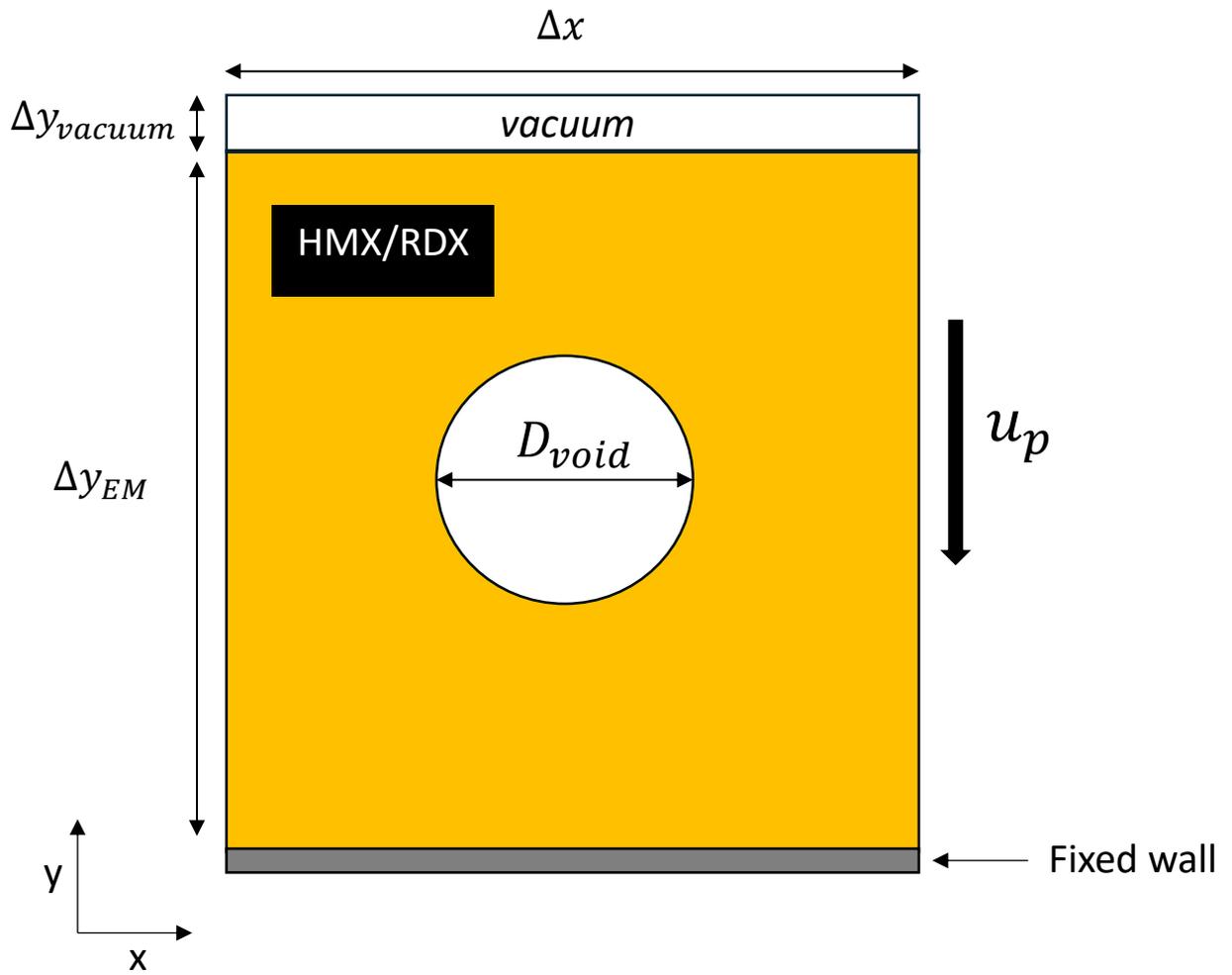

Figure 1. Schematic of physical setup for single pore collapse simulations. Range of pore diameters and particle speeds are described in Table 1.

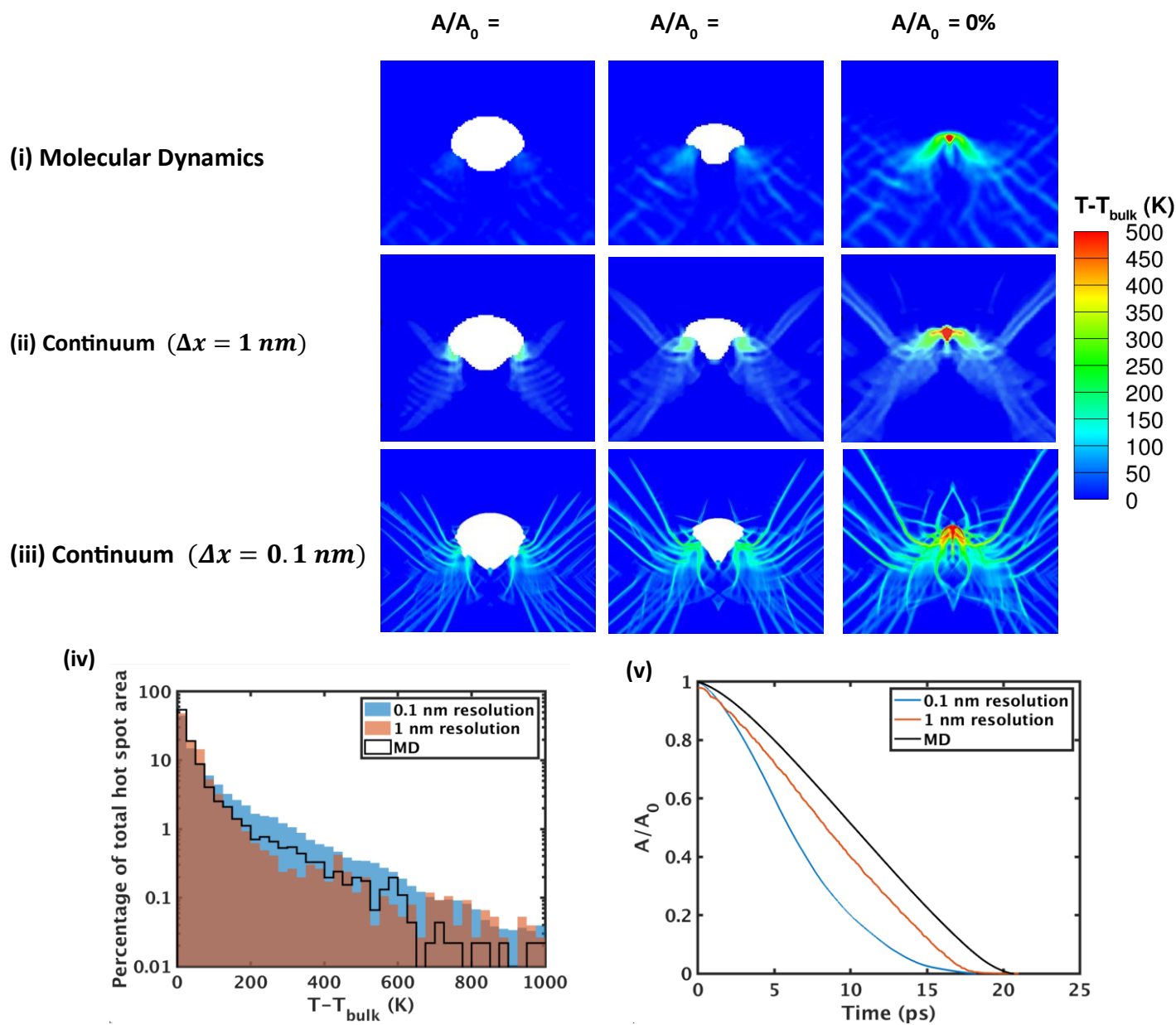

Figure 2. Comparison of temperature contours during collapse of 50 nm, circular pore in RDX with impact speed of 1 km/s. Showing instances when 50%, 25%, and 0% of the initial pore area is remaining. (i) MD simulation, (ii) Continuum simulation with grid resolution of $\Delta x = 1\ nm$, (iii) Continuum simulation with grid resolution of $\Delta x = 0.1\ nm$. (iv) Temperature distributions at instant of collapse and (v) pore collapse rate for MD and continuum simulations.

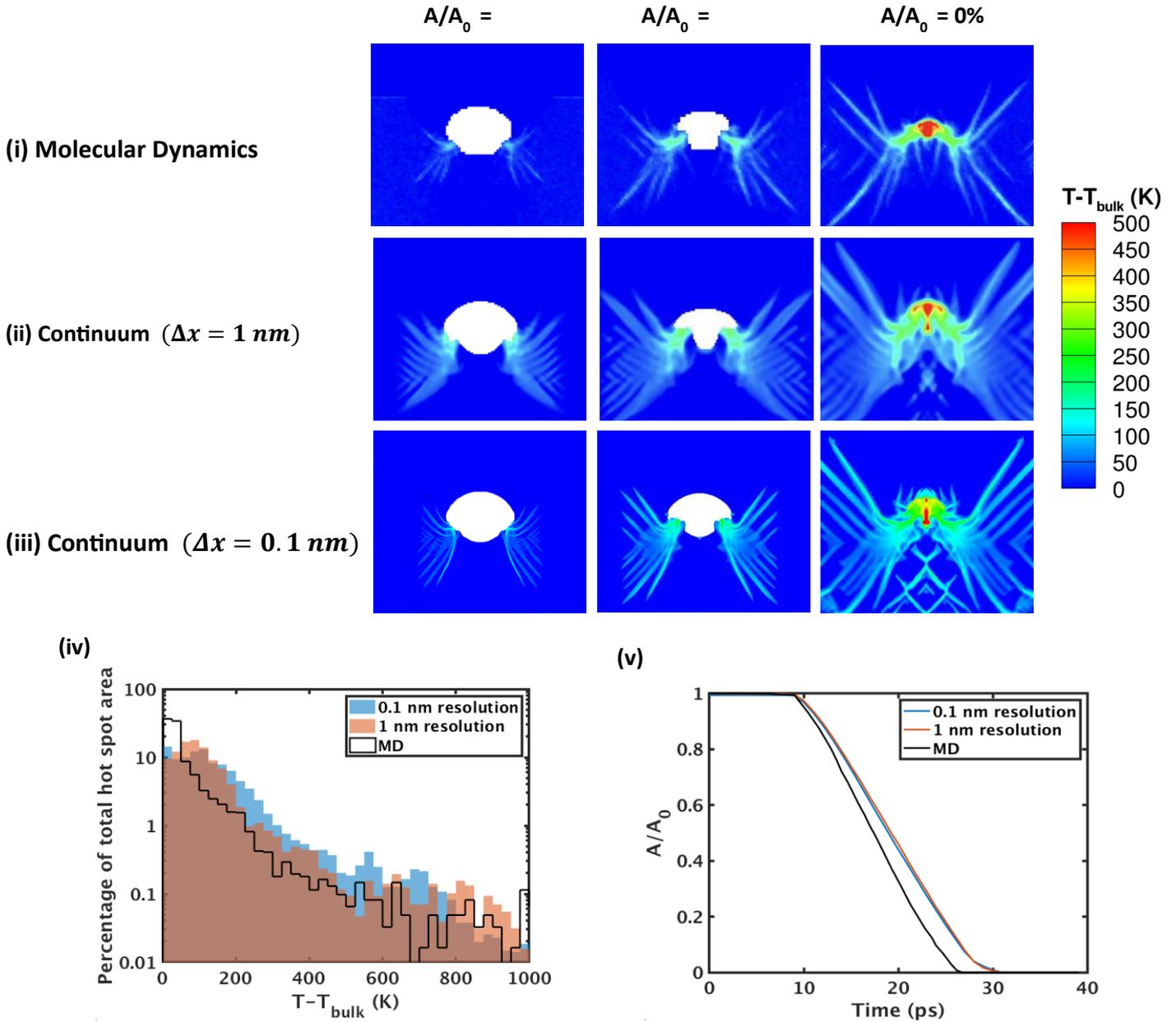

Figure 3. Comparison of temperature contours during collapse of 50 nm, circular pore in HMX. With impact speed of 1 km/s. Showing instances when 50%, 25%, and 0% of the initial pore area is remaining. (i) MD simulation, (ii) Continuum simulation with grid resolution of $\Delta x = 1\ nm$, (iii) Continuum simulation with grid resolution of $\Delta x = 0.1\ nm$. (iv) Temperature distributions at instant of collapse and (v) pore collapse rate for MD and continuum simulations.

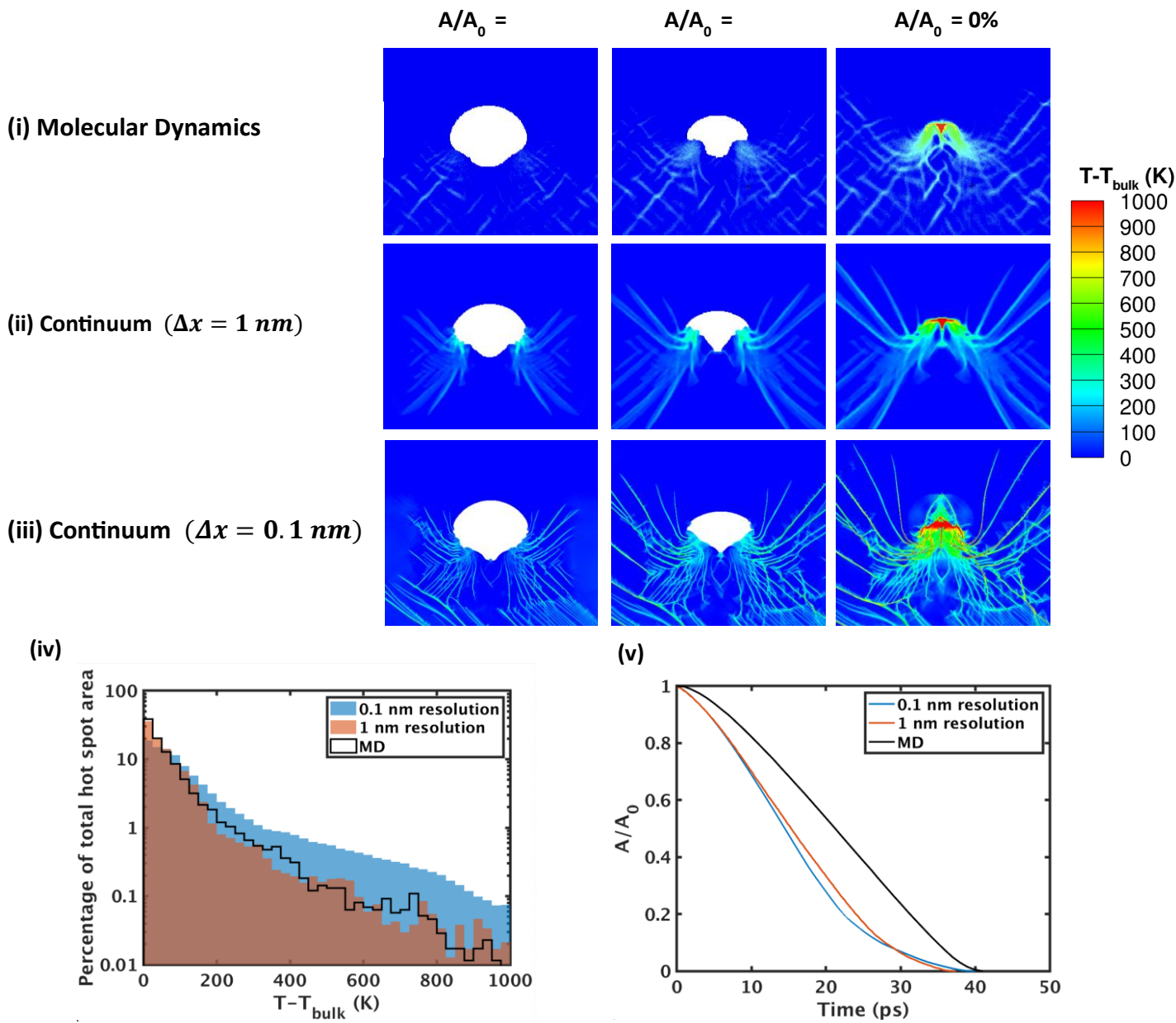

Figure 4. Comparison of temperature contours during collapse of 100 nm, circular pore in RDX with impact speed of 1 km/s. Showing instances when 50%, 25%, and 0% of the initial pore area is remaining. (i) MD simulation, (ii) Continuum simulation with grid resolution of $\Delta x = 1\ nm$, (iii) Continuum simulation with grid resolution of $\Delta x = 0.1\ nm$. (iv) Temperature distributions at instant of collapse and (v) pore collapse rate for MD and continuum simulations.

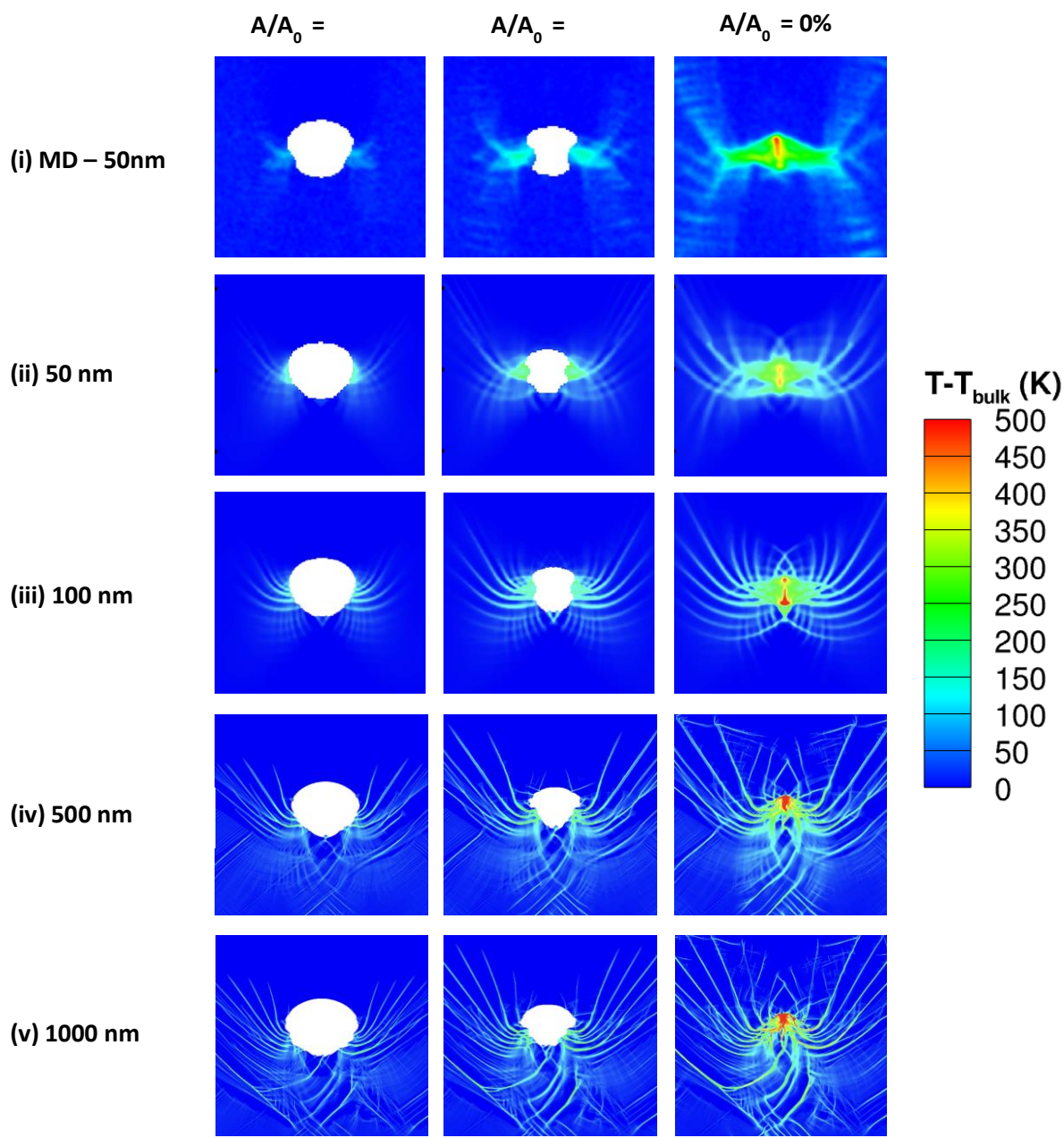

Figure 5. Temperature contours depicting pore and hot spot shapes during collapse of a single, circular pore in RDX with a particle speed of 0.5 km/s. (i) MD simulation of 50 nm pore. Also shown are continuum calculations with void diameters of (ii) 50 nm, (iii) 100 nm, (iv) 500 nm, and (v) 1000 nm. Each column represents different instances during collapse with 50%, 25%, and 0% of the initial pore area remaining.

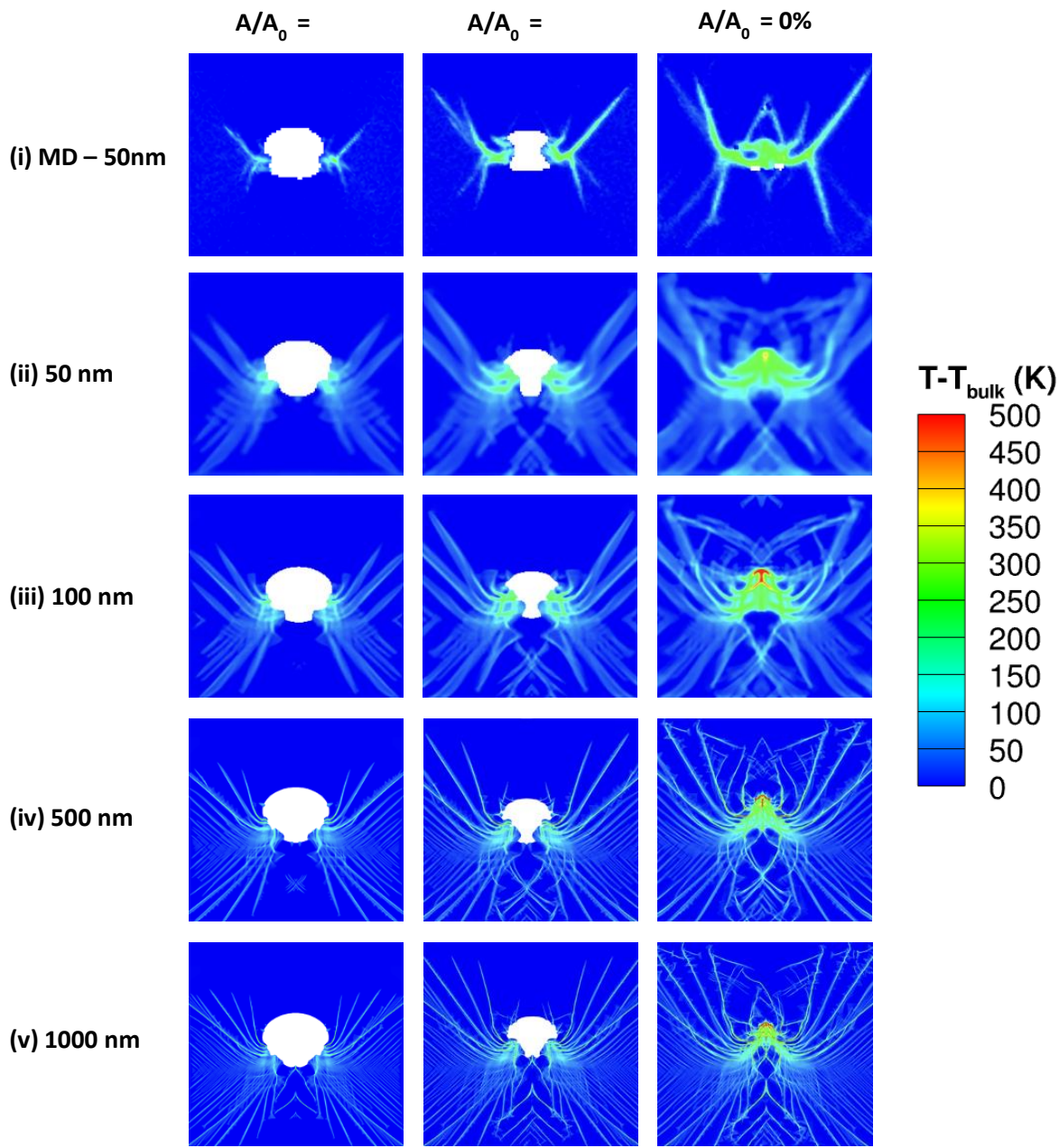

Figure 6. Temperature contours depicting pore and hot spot shapes during collapse of a single, circular pore in HMX with a particle speed of 0.5 km/s. (i) MD simulation of 50 nm pore. Also shown are continuum calculations with void diameters of (ii) 50 nm, (iii) 100 nm, (iv) 500 nm, and (v) 1000 nm. Each column represents different instances during collapse with 50%, 25%, and 0% of the initial pore area remaining.

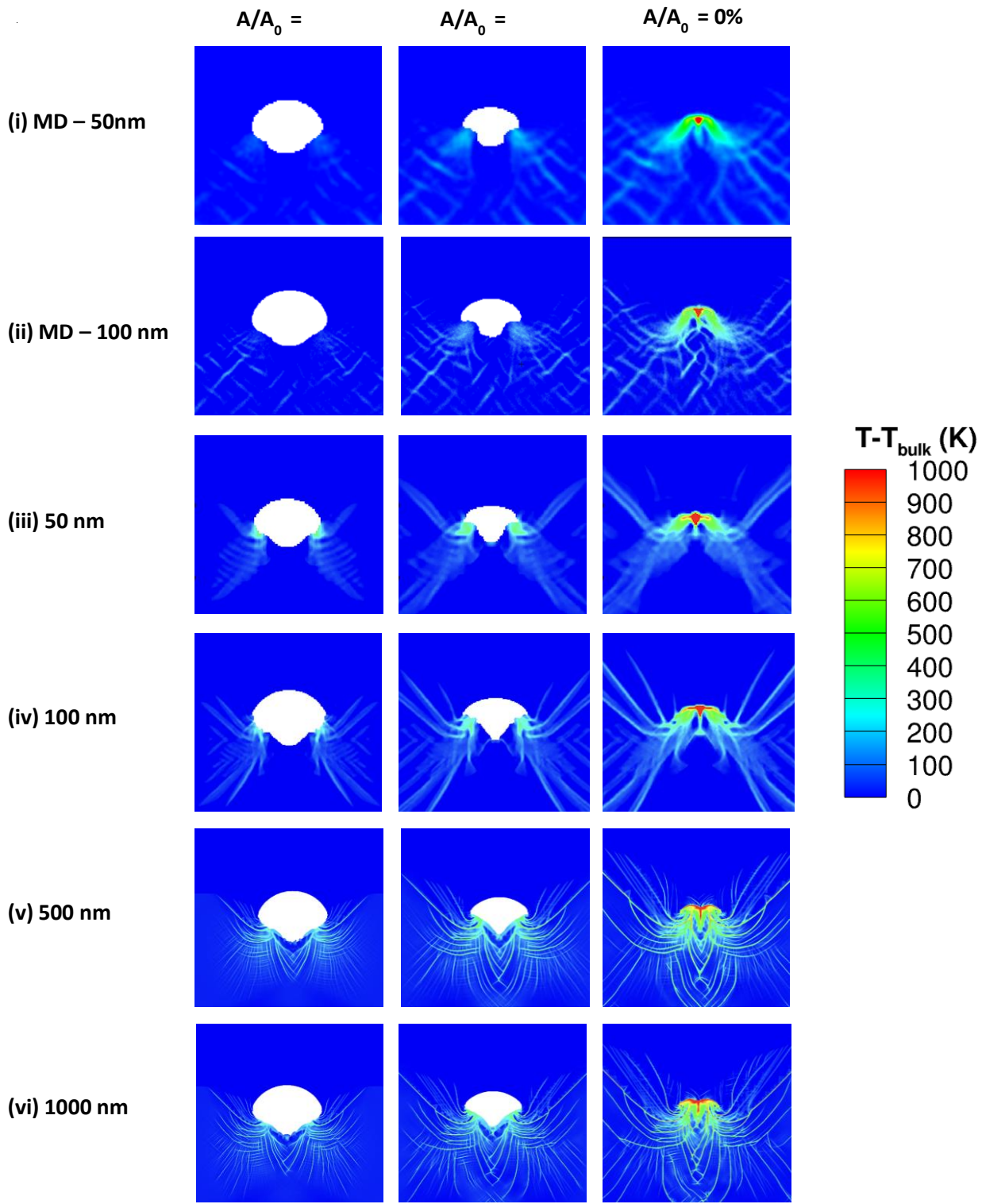

Figure 7. Temperature contours depicting pore and hot spot shapes during collapse of a single, circular pore in RDX with a particle speed of 1 km/s. MD simulations of: (i) 50 nm and (ii) 100 nm pore. Also shown are continuum calculations with void diameters of (iii) 50 nm, (iv) 100 nm, (v) 500 nm, and (vi) 1000 nm. Each column represents different instances during collapse with 50%, 25%, and 0% of the initial pore area remaining.

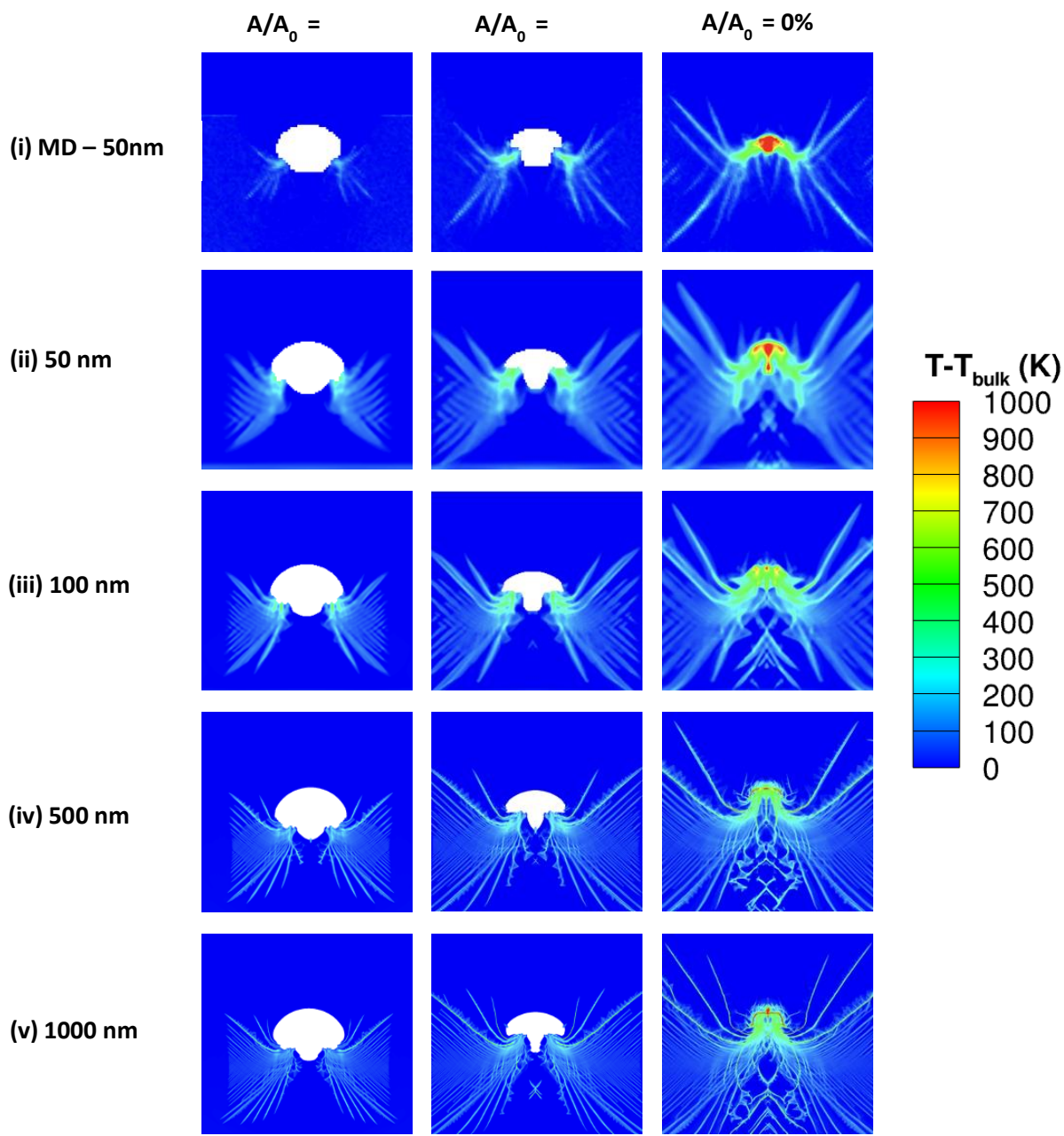

Figure 8. Temperature contours depicting pore and hot spot shapes during collapse of a single, circular pore in HMX with a particle speed of 1 km/s. (i) MD simulation of 50 nm pore. Also shown are continuum calculations with void diameters of (ii) 50 nm, (iii) 100 nm, (iv) 500 nm, and (v) 1000 nm. Each column represents different instances during collapse with 50%, 25%, and 0% of the initial pore area remaining.

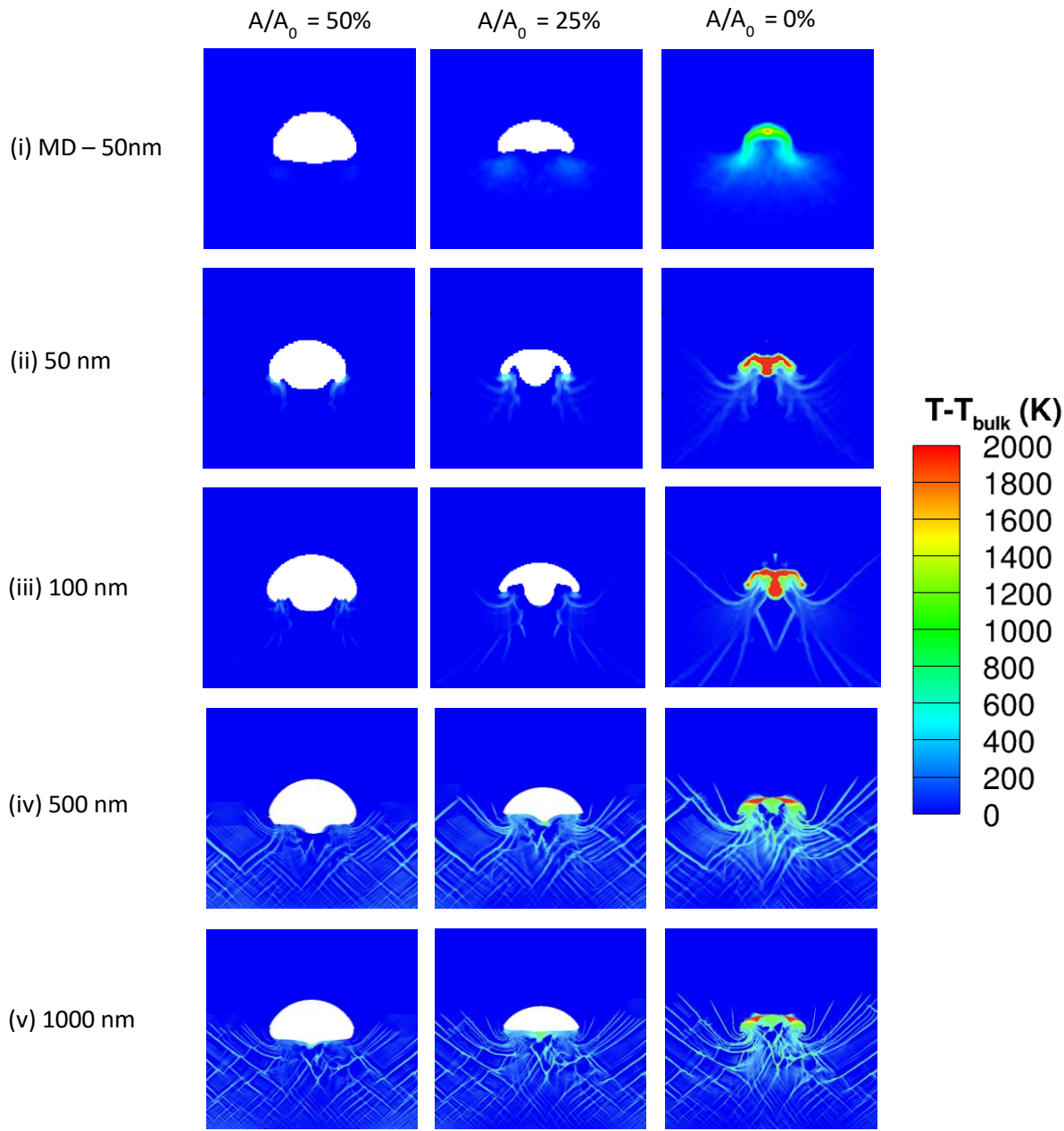

Figure 9. Temperature contours depicting pore and hot spot shapes during collapse of a single, circular pore in RDX with a particle speed of 2 km/s. (i) MD simulation of 50 nm pore. Also shown are continuum calculations with void diameters of (ii) 50 nm, (iii) 100 nm, (iv) 500 nm, and (v) 1000 nm. Each column represents different instances during collapse with 50%, 25%, and 0% of the initial pore area remaining.

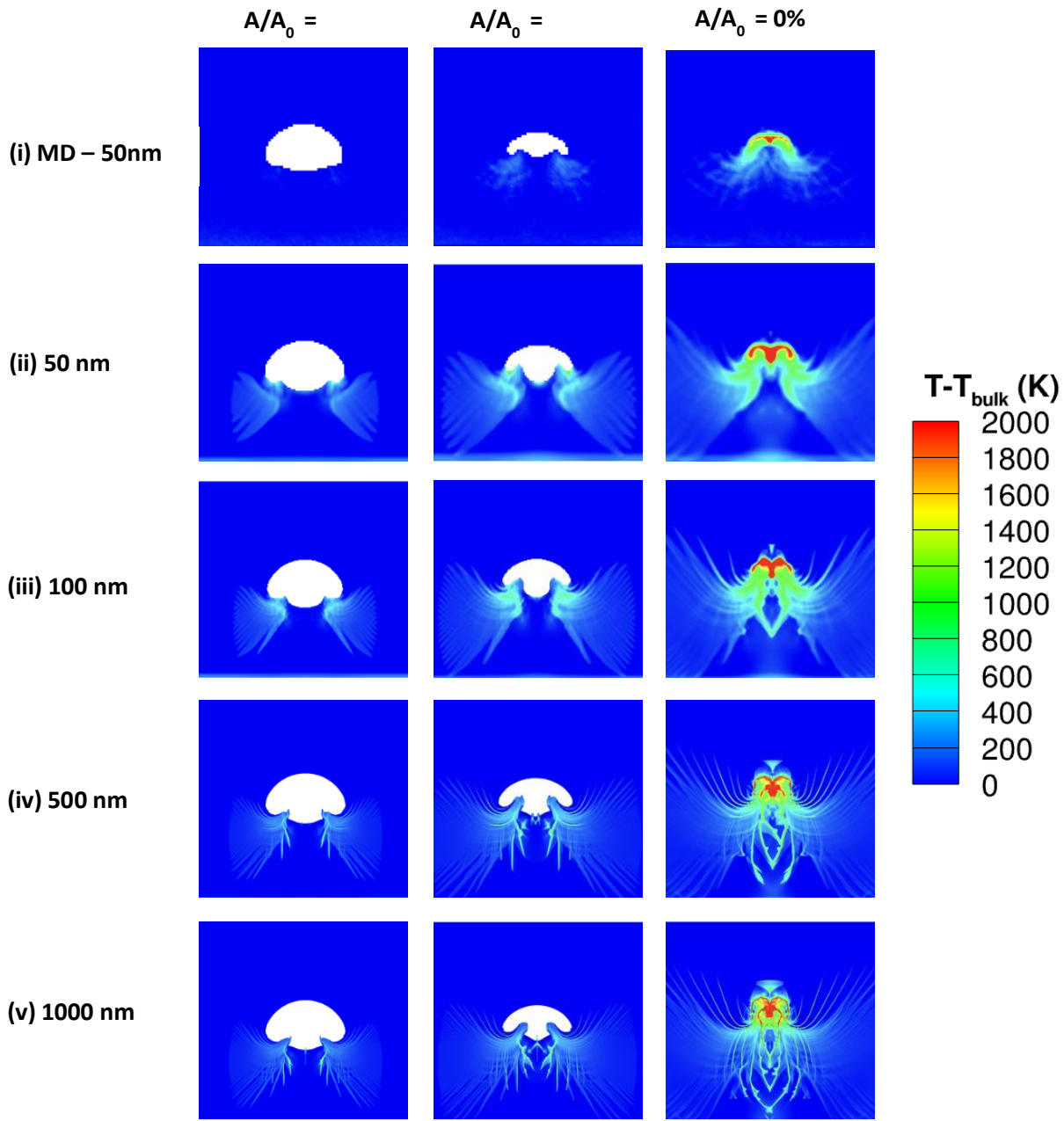

Figure 10. Temperature contours depicting pore and hot spot shapes during collapse of a single, circular pore in HMX with a particle speed of 2 km/s. (i) MD simulation of 50 nm pore. Also shown are continuum calculations with void diameters of (ii) 50 nm, (iii) 100 nm, (iv) 500 nm, and (v) 1000 nm. Each column represents different instances during collapse with 50%, 25%, and 0% of the initial pore area remaining.

(a)

**RDX** **HMX**

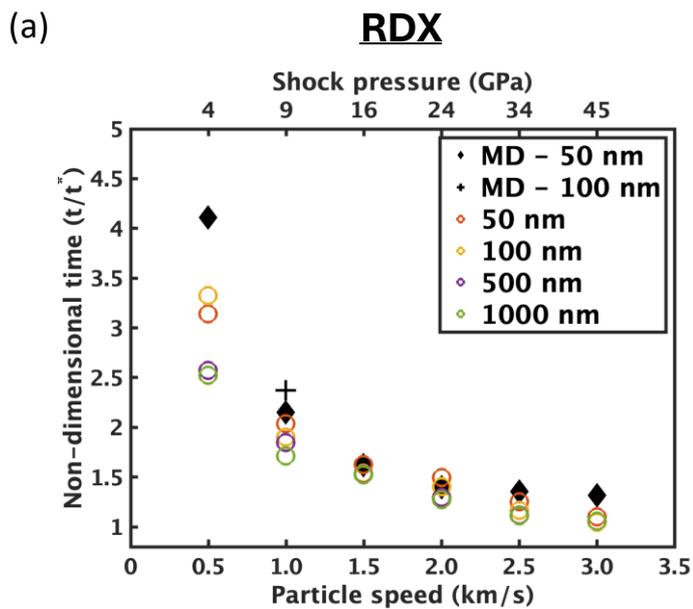
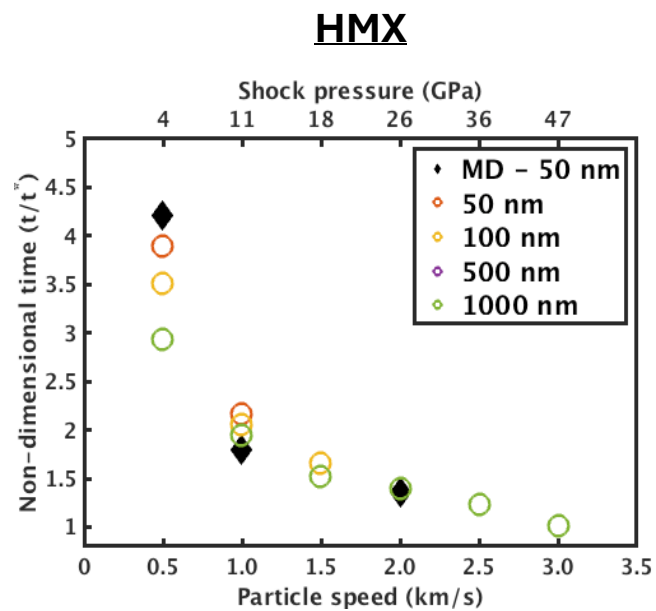

(b)

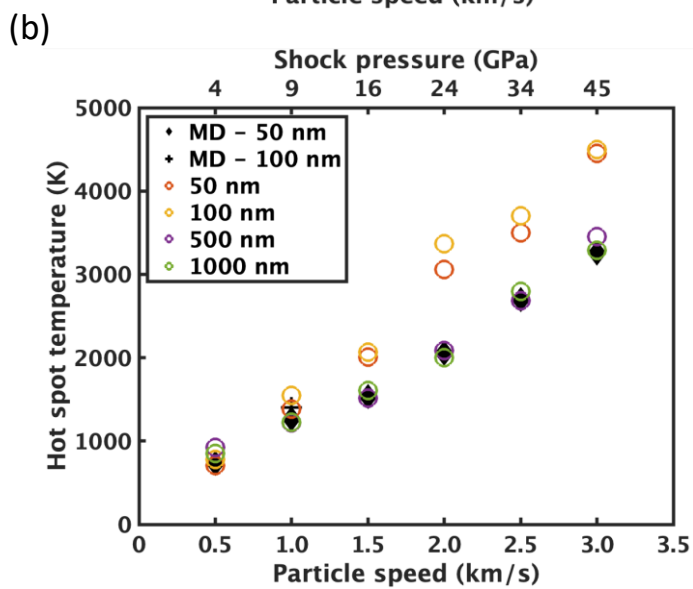
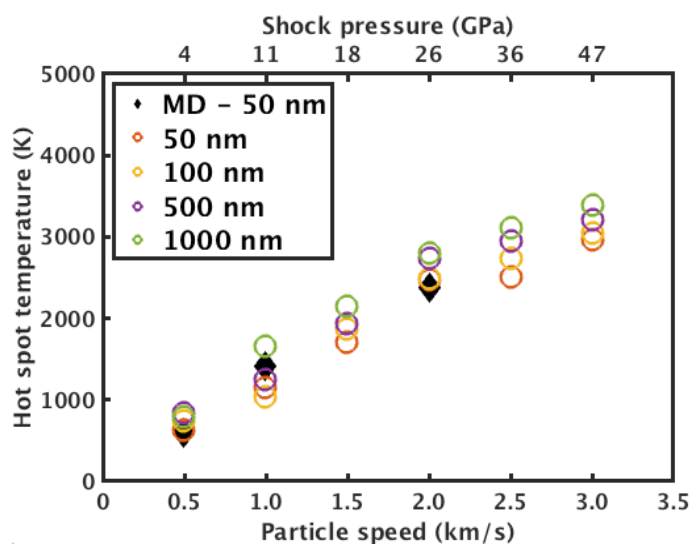

(c)

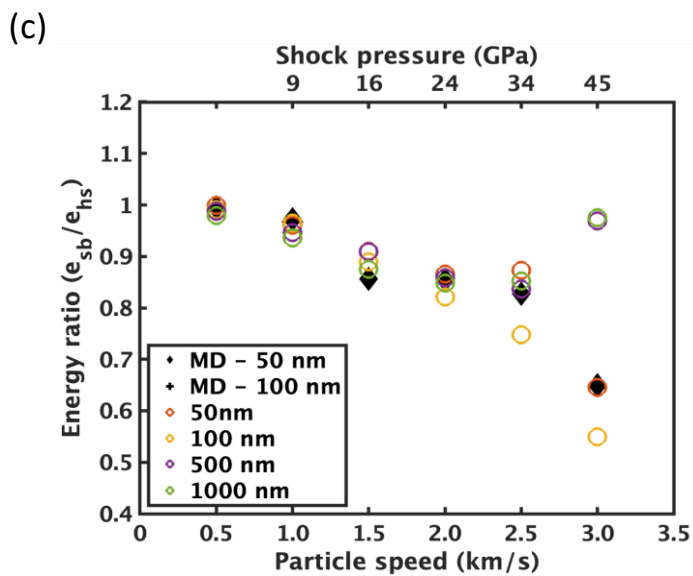
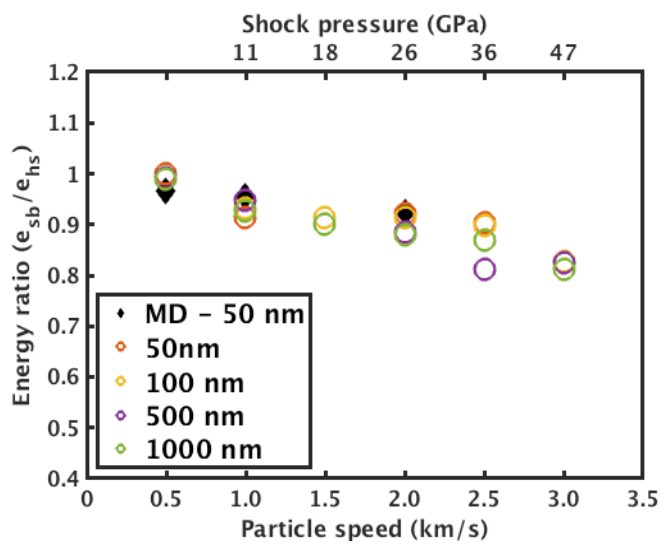

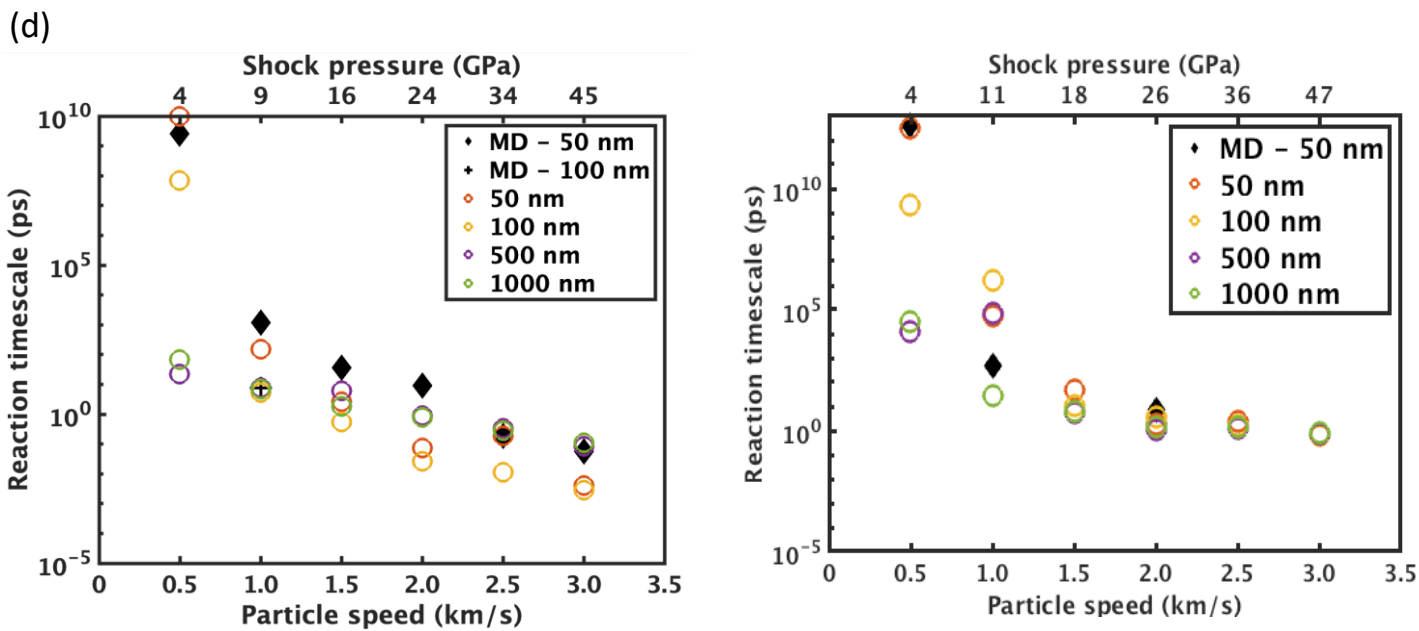

Figure 11. MD and Continuum with 1 nm grid resolution comparisons of pore collapse and hot spot metrics for all pore sizes and particle speeds. Left column shows RDX simulations, while right column shows HMX. Plots include (a) non-dimensionalized pore collapse duration, (b) average hot spot temperature, (c) ratio of shear band energy over total energy in hot spot, and (d) predicted timescale of reaction.

TABLE I. MD-consistent models and parameters used in the continuum calculations.

| Expression | RDX | Ref | HMX | Ref |
|---|---|---|---|---|
| $p_c(V) = \frac{3}{2} K_0 \left[ \left(\frac{V}{V_0}\right)^{-\frac{7}{3}} - \left(\frac{V}{V_0}\right)^{-\frac{5}{3}} \right]$ $* \left[ 1 + \frac{3}{4}(K_0' - 4)\left\{\left(\frac{V}{V_0}\right)^{-\frac{2}{3}} - 1\right\} \right]$ | $K_0 = 13\ GPa$ $K_0' = 9.2$ | [] | $K_0 = 16.3\ GPa$ $K_0' = 9.1$ | [] |
| $G_{RDX}(P,T) = G_0 + a_1 P + a_2(T - T_0)$ $G_{HMX}(P) = G_0 + a_1 P + a_2 P^2$ | $G_0 = 5.31\ GPa$ $a_1 = 3.3774$ $a_2 = -10.36e6$ | [] | $G_0 = 7.89\ GPa$ $a_1 = 1.7531$ $a_2 = -2.054e - 11$ | [] |
| $\Gamma = \Gamma_0 + \gamma_1 \left(\frac{\rho_0}{\rho}\right)^{\beta_1} + \gamma_2 \left(\frac{\rho_0}{\rho}\right)^{\beta_2}$ | $\Gamma_0 = 0.667$ $\gamma_1 = 2.01$ $\gamma_2 = -0.806$ $\beta_1 = 1$ $\beta_2 = 2$ | [] | $\Gamma_0 = 0.557$ $\gamma_1 = -1.976$ $\gamma_2 = 1.845$ $\beta_1 = 8.274$ $\beta_2 = 8.518$ | [] |
| $S^y = (A + B\epsilon_{ps}^n)\left[1 + C ln\left(\frac{\dot{\epsilon}_{ps}}{\dot{\epsilon}_0}\right)\right]\left[1 - \left(\frac{T - T_{ref}}{T_{melt} - T_{ref}}\right)^m\right]$ | $A = 0.3\ GPa$ $B = 0.1\ GPa$ $C = 1.819$ $m = 3$ $n = 0.1$ $\dot{\epsilon}_0 = 4.36e4\ s^{-1}$ | [] | $A = 0.1\ GPa$ $B = 0.1\ GPa$ $C = 2.172$ $m = 3$ $n = 0.1$ $\dot{\epsilon}_0 = 1e3\ s^{-1}$ | [] |
| $T_m(P) = T_{m,ref}\left[1 + \frac{P - P_{ref}}{a}\right]^{\frac{1}{c}}$ | $T_{m,ref} = 478\ K$ $P_{ref} = 0.0001\ GPa$ $a = 0.963\ GPa$ $c = 2.8855$ | [] | $T_{m,ref} = 551\ K$ $P_{ref} = 0\ GPa$ $a = 0.305\ GPa$ $c = 3.27$ | [] |
| $C_{v,RDX} = 1980\ J\ kg^{-1}\ K^{-1}$ $C_{v,HMX} = 2359\ J\ kg^{-1}\ K^{-1}$ | N/A | [] | N/A | [] |

TABLE II. Arrhenius law reaction parameters.

| Parameter | RDX | Ref. | HMX | Ref. |
|---|---|---|---|---|
| $E_a$ | $1.65e5\ \frac{J}{mol}$ | [] | $1.81e5\ \frac{J}{mol}$ | [] |
| $\ln(Z)$ | 38.66 | [] | 37.87 | [] |